\documentclass[fleqn,usenatbib]{mnras}

\usepackage{newtxtext,newtxmath}

\usepackage[T1]{fontenc}
\usepackage{color}

\usepackage{makecell}
\usepackage{threeparttable}

\DeclareRobustCommand{\VAN}[3]{#2}
\let\VANthebibliography\thebibliography
\def\thebibliography{\DeclareRobustCommand{\VAN}[3]{##3}\VANthebibliography}

\definecolor{bczhu}{rgb}{0,0.5,1}

\usepackage{graphicx}	
\usepackage{subfigure}
\usepackage{amsmath}	

\title[MACER-based AGN Modelling in TNG]{Towards physically more comprehensive AGN modelling in cosmological simulations: A MACER-based modification of IllustrisTNG}

\author[B. Zhu et al.]{%
Bocheng Zhu,$^{1,2,4}$
Volker Springel,$^{2}$\thanks{E-mail: vspringel@mpa-garching.mpg.de}
and Feng Yuan$^{3}$\thanks{E-mail: fyuan@fudan.edu.cn}%
\vspace{0.1cm}*\\%
$^{1}$Key Laboratory for Research in Galaxies and Cosmology, Shanghai Astronomical Observatory, Chinese Academy of Sciences, \\ 80 Nandan Road, Shanghai 200030, People's Republic of China\\%
$^{2}$Max-Planck-Institut für Astrophysik, Karl-Schwarzschild-Straße 1, 85741 Garching, Germany\\%
$^{3}$Center for Astronomy and Astrophysics and Department of Physics, Fudan University, 2005 Songhu Road, Shanghai, 200438,  People's Republic of China\\
$^{4}$Institute of Astrophysics, School of Physics, Zhengzhou University, China
}

\date{Accepted XXX. Received YYY; in original form ZZZ}

\pubyear{2015}

\begin{document}
\label{firstpage}
\pagerange{\pageref{firstpage}--\pageref{lastpage}}
\maketitle

\begin{abstract} 
Active galactic nuclei (AGN) feedback plays a significant role in many aspects of galaxy formation and evolution and has become a key ingredient in cosmological simulations. However, the subgrid models of AGN feedback in cosmological simulations such as IllustrisTNG (hereafter TNG) often overlook recent progress in the small-scale modelling of black hole (BH) accretion and AGN physics.  In this study, we improve on this by incorporating central aspects of the MACER model, a framework that treats AGN physics in greater detail, into the TNG feedback implementation. Specifically, we adopt MACER-prescriptions for feedback output for high and low accretion rates in a new model while the estimation of the accretion rate remains unchanged. We test this updated scenario both for idealized elliptical galaxies and for a cosmological box. Compared to the original TNG model, the MACER-based simulation leads to a higher star formation rate (SFR) and BH accretion rate in ellipticals, yielding a gas density profile in better agreement with observations. In the cosmological simulations, the time evolution of the SFR density, galaxy stellar mass function at $z=0$, and $M_{\star}-M_{\rm BH}$ relation at $M_{\star}>10^{10.5}\,{\rm M_{\odot}}$ are similar in both models. The MACER model better reproduces low-mass BHs in low-mass galaxies, and yields milder quenching in massive galaxies, although this is  accompanied by the absence of a pronounced colour bimodality. Still, the similarity of the outcomes underlines the self-regulated nature of BH feedback: for different feedback energetics, the accretion rate tends to adjust such that a similar total AGN feedback energy is released.

\end{abstract}




\section{Introduction}

Active galactic nucleus (AGN) feedback is believed to be a critical process in galaxy formation and evolution, responsible for many observed phenomena, such as the bimodality of galaxy colours \citep{strateva01,Kauffmann03,baldry04,bell04,faber07}, the low star formation rate (SFR) of massive galaxies \citep{salim07,noeske07,keroy08,Schiminovich2010}, the exponential decline of the galaxy stellar mass function (GSMF) at the massive end \citep{baldry08,baldry12,bernardi13}, and the scaling relations between supermassive black holes (SMBHs) and their host galaxies \citep{magorrian98,gultekin09,kormendy13}.

Over the past decades, numerous theoretical studies have incorporated AGN feedback into semi-analytic models \citep{croton2006,bower06,monaco2007,somerville2008}, galaxy-scale simulations \citep[e.g.,][]{springel05,dimatteo05,hopkins17,yuan18}, and cosmological simulations \citep[e.g.,][]{sijacki07,dubois14,choi15,crain15,weinberger17,dave19} in order to explain these observations. 

Depending on the black hole accretion rate relative to the Eddington rate, AGN feedback is commonly divided into two modes, referred to as radio mode and  quasar mode. The quasar mode corresponds to the high-accretion rate regime and was the earliest feedback mode investigated in simulations. \citet{dimatteo05} and \citet{springel05} implemented AGN feedback by injecting thermal energy from the AGN into the surrounding medium. They found that AGN feedback can help suppress star formation and black hole (BH) growth. Later, \citet{hopkins06} proposed a model of merger-triggered galaxy quenching and emphasized the role of AGN feedback in this process. More recently, \citet{davies2022,davies24} studied the impact of AGN feedback on SMBHs and their host galaxies during mergers by using the genetic modification technique to alter the merger histories of Milky Way–sized galaxies, and found that SMBH and galaxy properties change significantly owing to the nonlinear nature of mergers. In the quasar mode, radiation is considered to play a particularly significant role. For example, \citet{costa18b} performed a cosmological zoom-in simulation with radiative transfer and found that the radiation pressure on the dust can efficiently trigger galactic outflow, with \citet{barnes20} reaching similar conclusions.

It is clear, however, that a black hole spends most of the time in the low-accretion state, the so-called radio mode. After early studies pointed out an important feedback impact of the radio mode \citep{croton2006,sijacki07}, it has since developed into a key research area, especially for massive galaxies and galaxy clusters \citep[e.g.,][]{gaspari12,li15,weinberger17a,su21,2025arXiv251102796H}. In a seminal analysis of the Millennium simulation, \citet{croton2006} has incorporated  radio mode AGN feedback in their semi-analytic model of galaxy formation and found that it can successfully reproduce the exponential cutoff of the galaxy stellar mass function (GSMF) at the high mass end, thereby reproducing the old stellar ages of massive galaxies. Later, \citet{sijacki07} proposed a numerical subgrid model for radio mode feedback in hydrodynamical simulations. They injected thermally heated bubbles  to mimic the bipolar energy injection of jets and the bubbles they inflate in galaxy clusters. \citet{gaspari12} have likewise kept massive galaxies in the quiescent state by incorporating mechanical feedback from AGN.  \citet{li15} and \citet{qiu20} have incorporated kinetic AGN jets into the simulation of isolated galaxy clusters and found that the AGN feedback can effectively suppress cooling flows and reproduce a mass and morphology of the HI gas consistent with observations. 

More recently, \citet{weinberger17} proposed a kinetic feedback model for the radio mode and implemented it in the IllustrisTNG project. This model successfully reproduces the quenching of star-formation in large galaxies, while keeping gas fractions comparatively high. The latter is actually not favoured any more by most recent data, such as eROSITA gas fractions \citep{Popesso2024} or observations of the kinetic SZ effect \citep{Hadzhiyska2025}, which both suggest lower gas fractions and thus even stronger AGN feedback than in TNG. 

\citet{zhu23} have studied the role of AGN feedback in the late-stage evolution of a massive elliptical galaxy based on the galactic-scale high resolution AGN feedback model MACER \citep{yuan18}.  Different from some previous results \citep[e.g.,][]{gaspari12}, they found that although the AGN spends most of its time in the radio (hot) mode, the wind feedback in the quasar mode actually dominates the cumulative energy output of AGN and plays the dominant role in the galaxy evolution.  

Despite the large associated uncertainties, AGN feedback is now widely accepted as a fundamental ingredient in cosmological simulations of galaxy formation \citep[][etc.]{dubois14, vogelsberger14, crain15, schaye15, weinberger17, dave19, schaye25}. Only with the inclusion of AGN feedback, these simulations can successfully reproduce principal observational results for whole galaxy populations, such as the GSMF, the bimodality of the galaxy colour distribution, the $M_{\rm BH}-M_{\rm \star}$ relation, etc. However, the current modelling of AGN in these simulations is usually done in a heuristic and parameterized way, where the parameters are freely tuned to best match the observational results. It would therefore be highly desirable to improve the physical grounding of these models.

Over the past few years, several groups have made efforts to develop AGN feedback models with stronger physical motivation. \citet{bustamante2019} incorporated black hole spin evolution with a spin-dependent radiative efficiency, while \citet{dubois2021} further included spin-dependent jet efficiency based on GRMHD simulations in cosmological (zoom) simulations. \citet{farcy2025} introduced a kinetic wind model for radiatively efficient accretion, motivated by observations of AGN disk winds \citep{tombesi2013} and building upon early simulation studies of AGN wind impacts on galaxy properties \citep{ostriker10,choi2012}. More recently, \citet{husko2024,husko2025} presented a comprehensive hybrid AGN feedback model and applied it to the COLIBRE simulations, which incorporates distinct accretion disk states (thick, thin, and slim), tracks black hole spin evolution, and self-consistently drives both thermal and kinetic feedback depending on the accretion state.

An important aspect that has received increasing attention in recent years is the role of winds in the radio mode. In previous numerical simulations, jets were often considered to be the main AGN feedback output in the radio mode. However, both theoretical \citep{yuan12b,narayan12,yuan15,yang21} and observational \citep{cheung16,shi21,2022ApJ...926..209S} studies have shown in recent years that winds and jets are symbiotic and act together in  hot accretion flows around black holes. Moreover, detailed three-dimensional general relativistic MHD simulations of black hole accretion have shown that, even in the extreme case of a most rapidly spinning black hole and a strongly magnetized accretion flow \cite[i.e., a magnetically arrested accretion disk,][]{2003PASJ...55L..69N}, the momentum flux of winds is larger than that of jets, although the energy flux of the winds is weaker \citep{yuan15,yang2021b}. This result, combined with the fact that the opening angle of a wind is much larger than that of a jet, suggests that winds may play a more important role in the radio mode AGN feedback than commonly assumed. 

In recent years, the importance of winds in the radio mode has begun to be taken seriously \citep{weinberger17, yoon19}. For example, \citet{yoon19} find that without the wind in the radio feedback mode, the ``typical'' AGN luminosity would be significantly higher, and the black hole growth would be larger. And interestingly, these theoretical results have been met with strong observational support. For example, X-ray observations of M81* and NGC 7213 have detected that winds are shock heating and impeding the interstellar medium surrounding the black hole \citep{2024ApJ...970...48S}. Most recently, \citet{2025arXiv251102796H} have incorporated both jets and winds in a fully self-consistent manner in a study of the canonical cooling flow problem in galaxy clusters, with  parameters being determined by general relativistic magnetohydrodynamic simulations of black hole accretion and by observations. They found that the jet-wind shear induces strong turbulence, whose dissipation efficiently converts the kinetic energy of the jet into thermal energy of the surrounding gas. This mechanism allows their model to successfully reproduce key observables -- including the cold-gas mass, star formation rate, radial thermodynamic profiles of the cluster gas, and black hole growth. 

Nevertheless, these aspects of radio mode feedback physics have so far not been widely incorporated in numerical studies of AGN feedback. A notable exception is the MACER  model \citep{yuan18, 2018ApJ...864....6Y,2025ApJ...985..178Z}. The key features of MACER are as follows. First, as it concentrates on galaxy rather than cosmological scales, it can afford a much higher spatial resolution. As the inner boundary of the simulation domain in MACER is smaller than the outer boundary of the black hole accretion flow, i.e., the Bondi radius, the mass flux can be  directly measured at the inner boundary of the simulation domain. Combining this with the theory of black hole accretion, the expected mass accretion rate at the black hole horizon can then be reliably assessed. Note that this quantity is the most crucial parameter for the study of AGN feedback since it determines the power of the AGN. Secondly, a comprehensive state-of-the-art physics description of black hole accretion has been adopted in MACER, i.e., the radiation and outflow characteristics of the AGN at low and high mass accretion rates are treated in detail. The MACER model has successfully reproduced many observations of massive elliptical galaxies such as the AGN duty cycle and X-ray luminosity \citep{yuan18,2018ApJ...864....6Y}.  In further applications, the model has been used to investigate the black hole fueling and feedback in a compact galaxy \citep{2023MNRAS.523.1641D}, the impact and fate of cosmological inflow in the context of elliptical galaxy \citep{2023MNRAS.524.5787Z}, the roles of different feedback modes (radio and quasar) and components (AGN radiation and outflow) \citep{zhu23}, the positive feedback in a starburst dwarf galaxy \citep{2025arXiv251020897S}, the cooling flow problem in galaxy clusters \citep{2025arXiv251102796H}, and the AGN feedback in a disk galaxy \citep{zou26a,zou26b}. 

Like these recent efforts, our implementation of basic-MACER in cosmological simulations aims to strengthen the physical grounding of AGN feedback, but with a distinct emphasis on directly adopting prescriptions from GRMHD simulations and observational constraints without additional calibration. Similar to recent  works, we distinguish between different accretion states, with a transition at $\dot{M}_{\rm BH}/\dot{M}_{\rm Edd} \simeq 0.01$ motivated by observations \citep{yuan14}. However, unlike the spin-dependent jet efficiency in \citet{husko2025} or magnetically chocked accretion discs-based jet model in \citet{dubois2021}, our approach directly adopts wind properties from GRMHD simulations \citep{yuan15} and ultra-fast object (UFO) observations \citep{gofford15} for hot and cold accretion modes, respectively. Compared to MISTRAL, which focuses on kinetic winds for radiatively efficient accretion, basic-MACER also incorporates winds in the low-accretion hot mode, where both theoretical \citep{narayan12,yuan15,yang21} and observational \citep{cheung16,shi21} studies indicate they play a significant role alongside jets. Another key distinction of our approach is the accretion fraction, which is defined as the ratio of BH accretion rate (the accretion rate on the BH horizon, hereafter BHAR) and Bondi accretion rate, that varies with accretion rate in the hot mode, reflecting mass loss through winds as predicted by GRMHD simulations and the change of the ``transition radius'' with accretion rate \citep{yuan18}, which is the feature also present in \citet{husko2025} but absent in most other models that assume a fixed accretion efficiency.

In this work, we now aim to  take a first step towards bridging the galactic-scale MACER model and cosmological simulations. Our main contributions are:
\begin{itemize}
    \item Implementation of a simplified version of the MACER model (hereafter basic-MACER) into the IllustrisTNG framework;
    \item Comparison of basic-MACER with the original AGN model in TNG both in an idealized elliptical galaxy and in cosmological box runs;
    \item Quantification of their differences in star formation, black hole growth, and baryonic properties;
    \item Discussion of the physical implications and limitations of incorporating more realistic AGN physics into large-scale cosmological simulations.
\end{itemize}

It is clear from the outset that the spatial resolution limitations of cosmological simulations are a serious obstacle in improving the AGN treatment, because the Bondi radius remains  typically  unresolved. As a first step, we therefore concentrate on the feedback description in the present work, leaving the accretion rate estimate unchanged. The latter could be addressed in future work by exploiting adaptive mesh resolution around the BH. Specifically, in the present paper we simply replace the AGN feedback model of
the cosmological IllustrisTNG project \citep[hereafter TNG;][]{springel18,naiman18,marinacci18,nelson18,pillepich18b}  {\it partly}  with the one used in MACER. In practice, we adopt the AGN output as a function of the mass accretion rate as computed in the MACER model, but other aspects of the AGN physics models remain unchanged compared to the TNG project. In this way, we seek to assess how robust the outcomes are with respect to a strong change in the way the energetics of AGN feedback are determined.

This paper is organized as follows. In Section~\ref{sec:model}, we introduce the black hole accretion and AGN feedback model in MACER and our approach of using it in cosmological simulations (we call it the ``basic-MACER'' model). The original AGN model adopted in TNG is also briefly summarized. We then present in Section~\ref{sec:setup} the initial setup for the idealized elliptical galaxy and cosmological simulation box we use to assess the models. In Section~\ref{sec:idealized}, we present the results of the basic-MACER and TNG models for the isolated elliptical galaxy simulations, while in Section~\ref{cosmological} we compare and discuss the performance of the basic-MACER and the TNG model in cosmological simulations. In Section~\ref{sec:param}, we assess the parameter and numerical resolution dependence of the AGN feedback treatment when using the basic-MACER model. We also discuss differences between the basic-MACER and the original high-resolution MACER model in this section. Finally, we conclude and summarize our results in Section~\ref{sec:conclusion}.

\section{Models}\label{sec:model}

In this section, we describe the implementation of black hole physics in the basic-MACER model, which builds upon the MACER project. We also briefly summarize the black hole treatment in TNG. 

\subsection{Black hole seeding and growth}\label{sec:seed}

The basic-MACER model shares the black hole seeding, growth and numerical implementation of kinetic energy injection with the TNG model. Before we introduce both models, we briefly summarize the corresponding implementations. Please refer to \citet{weinberger17} for more details.

Seed black holes are placed in friend-of-friend (FoF) identified halos more massive than $5\times10^{10}\,{\rm M}_\odot$, with an initial mass of $M_{\rm seed}=1.18\times10^6\,{\rm M}_\odot$. Black holes grow via mergers and gas accretion. The accretion rate is estimated using the Bondi–Hoyle–Lyttleton prescription without a boost factor,
\begin{equation}\label{eq:mdotest}
\dot{M}_{\rm Bondi} = \min\left(\dot{M}_{\rm Bondi,\, std}, \dot{M}_{\rm Edd}\right),
\end{equation}
where
\begin{equation}\label{eq:bondiandedd}
\dot{M}_{\rm Bondi,\,std} = \frac{4\pi G^2 M_{\rm BH}^2\bar{\rho}}{\bar{c_s}^3}, \quad
\dot{M}_{\rm Edd} = \frac{4\pi G M_{\rm BH} m_p}{\epsilon_r \sigma_T c}.
\end{equation}
Here, $\bar{\rho}$ and $\bar{c_s}$ are kernel-weighted estimates of the local gas density and sound speed, respectively, and $\epsilon_r$ is the radiative efficiency assumed to be 0.2 for the TNG model and 0.1 for the basic-MACER model. In the magnetohydrodynamical TNG simulations, the effective sound speed includes magnetic pressure,
\begin{equation}
c_s = \left(c_{s,\rm therm}^2 + \frac{B^2}{4\pi \rho}\right)^{1/2}.
\end{equation}
The number of neighboring cells used in the kernel estimate is usually adjusted with resolution. We have set it to 48 in the simulations analyzed here.

\subsection{Kinetic feedback implementation}\label{sec:kfbi}

The TNG model adopts an empirical model for kinetic feedback since  winds or jets are largely unresolved in cosmological simulations. The kinetic feedback energy $\Delta E$ is distributed to neighboring cells within the BH feedback region (same as for accretion) by 
imparting momentum to them. A cell $j$ receives a momentum
\begin{equation}
\Delta \mathbf{p}_j = m_j \sqrt{\frac{2\Delta E\,w(\mathbf{r}_j)}{\rho}}\,\mathbf{n},
\end{equation}
where $m_j$ is the gas cell mass, $w(\mathbf{r}_j)$ is the kernel weight, $\rho$ is the local kernel-weighted density, and $\mathbf{n}$ is a unit vector for a randomly chosen direction (changed every timestep). Total momentum is therefore not strictly conserved in a single injection event, but the expectation value of the total added momentum vanishes when averaged over time.

To ensure effective outflows, a minimum energy threshold is imposed:
\begin{equation}
E_{\rm inj,min} = f_{\rm re}\,\tfrac{1}{2}\sigma_{\rm DM}^2\,m_{\rm enc},
\end{equation}
where $\sigma_{\rm DM}$ is the local dark matter velocity dispersion, $m_{\rm enc}$ is the enclosed gas mass within the feedback region, and $f_{\rm re}$ is a free parameter controlling burstiness (set to 1 in the default model). If the feedback energy is below this threshold it is accumulated in the BH particle until a sufficient amount is reached, which is then released.

\subsection{The AGN model in basic-MACER}

The full MACER model requires resolving the Bondi radius in order to self-consistently connect large-scale inflows to black hole accretion. Achieving this resolution in cosmological simulations is still computationally prohibitive. Therefore, in our basic-MACER implementation, the accretion rate estimation and the AGN–ISM coupling still follows the empirical treatment of TNG (see Section~\ref{TNG}), while the subgrid physics of black hole feedback is taken from MACER. We refer to this hybrid approach as ``basic-MACER''. A fully comprehensive description of the original MACER framework can be found in \citet{yuan18}.

\subsubsection{Accretion and feedback modes}

Accretion is divided into cold and hot modes, depending on the Eddington-scaled accretion rate or equivalently the AGN luminosity. Theoretical and observational studies \citep{mc06,yuan14} suggest a transition AGN luminosity $L_{\rm C}$ at
\begin{equation}
L_{\rm C} \sim 0.01\,L_{\rm Edd}.
\end{equation}
At high accretion rates, the accretion flow corresponds to a geometrically thin, optically thick disk, while at low accretion rates it transitions to a geometrically thick, optically thin hot flow. The two regimes differ fundamentally in their dynamics and energetics, leading to distinct scaling relations for radiation, wind, and jet outputs \citep{yuan18}. Accordingly, MACER distinguishes in its feedback between a cold (quasar) mode and a hot (radio) mode. 

The AGN luminosity is computed as
$L_{\rm AGN} = \epsilon \dot{M}_{\rm BH} c^2$,
where $\dot{M}_{\rm BH}$ is the black hole accretion rate (BHAR), which is the mass accretion rate on the BH horizon. However, in cosmological simulations, the Bondi radius can usually not be resolved. Instead the Bondi accretion rate is estimated. For simplicity, we use the Bondi accretion rate to separate the two modes, assuming that the Bondi rate and BHAR are comparable in the cold mode. Specifically, we classify the accretion as cold (quasar) mode when $\dot{M}_{\rm Bondi} > \dot{M}_{\rm crit}\equiv0.01\dot{M}_{\rm Edd}$, and as hot (radio) mode otherwise. This is regarded as a phenomenological assumption, to be refined in future work.

\subsubsection{Cold mode}
For the cold mode, both wind and radiation contribute to feedback in the original MACER model. However, previous works \citep{yoon19, zhu23} indicated that winds are the dominant feedback channel, especially in regulating the AGN luminosity and suppressing star formation. For simplicity, we thus only include winds in basic-MACER. Their mass outflow rate and velocity follow the empirical fits of \citet{gofford15}, who analyzed 51 AGNs with Suzaku and constrained UFOs at $\sim10^{2-4} r_{\rm s}$:
\begin{equation}\label{coldwindflux}
\dot{M}_{\rm wind,\,c} = 0.28 \left(\frac{L_{\rm bol}}{10^{45}\,\rm erg\,s^{-1}}\right)^{0.85} {\rm M}_\odot\,{\rm yr}^{-1},
\end{equation}
\begin{equation}
v_{\rm wind,\,c} = \min\left[2.5\times10^4\left(\frac{L_{\rm bol}}{10^{45}\,\rm erg\,s^{-1}}\right)^{0.4},\,10^5\right] {\rm km\,s^{-1}},
\end{equation}
where $\dot{M}_{\rm wind,\,c}$ is the wind mass outflow rate of in the cold mode, $v_{\rm wind,\,c}$ is the wind velocity in the cold mode, and $L_{\rm bol}\equiv\epsilon_{\rm r}\dot{M}_{\rm BH}c^2$ is the bolometric luminosity with $\epsilon_{\rm r}=0.1$. The fitting formula follows \citet{yuan18}, whose best fitting parameters are presented in Table 4 of \citet{gofford15}. The corresponding energy output $\dot{E}_{\rm cold}$ is
\begin{equation}\label{colde}
\dot{E}_{\rm cold} = \tfrac{1}{2}\dot{M}_{\rm wind,\,c}v_{\rm wind,\,c}^2.
\end{equation}
Numerical simulations of disk winds suggest that roughly half of the inflowing mass is lost again through outflows \citep[e.g.,][]{2019MNRAS.490.2567M}. 
We adopt a constant value of $f_{\rm acc}=0.5$ for the accretion fraction $f\equiv\dot{M}_{\rm{BH}}/\dot{M}_{\rm{Bondi}}$, to account for the mass loss due to disk winds.

\subsubsection{Hot mode}

The feedback outputs of accretion in the hot mode include winds, jets, and radiation. Similar to the cold mode, we ignore the radiation output in the hot mode since winds are more important \citep{yoon19}.  Although there is growing observational evidence for winds in low-luminosity AGNs \citep[e.g.,][]{cheung16, shi21}, the wind properties are still poorly constrained from observations. In contrast to the observations, GRMHD simulations of black hole accretion have systematically studied the launching and the properties of winds  \citep{yuan12a,yuan12b, narayan12, yuan15,yang21}. 

The geometry of the accretion flow in the hot mode is an outer truncated thin disk plus an inner hot accretion flow \citep{yuan14}, where the thin disk is truncated when radiative cooling becomes inefficient and the accretion flow transitions into a hot, optically thin mode due to evaporation or other processes \citep[e.g.,][]{liu1999, gu00, manmoto00, yuan04, yuan14}. The transition radius $r_{\rm tr}$ between them is determined by
\begin{equation}\label{rtr}
    r_{\rm{tr}}=3r_{\rm{s}} \left(\frac{2\times10^{-2}\dot{M}_{\rm{Edd}}}{\dot{M}_{\rm{Bondi}}}\right)^2.
\end{equation}
Here $\dot{M}_{\rm Bondi}$ is the Bondi accretion rate from the accretion rate estimation procedure, and $r_{\rm s}$ is the Schwarzschild radius of the black hole. This expression is modified from Eq. 18 of \citet{yuan18}, whose original expression of the transition radius is parameterized as $r_{\rm tr}=r_{\rm tr}\left(\dot{M}(r_{\rm in})\right)$, where $\dot{M}(r_{\rm in})$ is the mass inflow rate measured at the inner boundary of their simulation. In our implementation, we approximate the mass inflow rate by the Bondi accretion rate, i.e., $\dot{M}(r_{\rm in})\approx\dot{M}_{\rm Bondi}$. Therefore, Eq.~(\ref{rtr}) is a reformulation of the model of \citet{yuan18} expressed in terms of the Bondi accretion rate used in our subgrid prescription.

The accretion flow is assumed to not exhibit mass loss outside of the transition radius. Following \citet{yuan15}, the mass outflow rate $\dot{M}_{\rm{wind,\,h}}$ and the wind velocity $v_{\rm{wind,\,h}}$ in hot mode are formulated as
\begin{equation}\label{mdoth}
    \dot{M}_{\rm{wind,\,h}} = \dot{M}_{\rm{inflow}} (r_{\rm tr}) \left[1-\left(\frac{3r_{\rm{s}}}{r_{\rm tr}}\right)^{0.5}\right]\approx \dot{M}_{\rm Bondi} \left[1-\left(\frac{3r_{\rm{s}}}{r_{\rm tr}}\right)^{0.5}\right],
\end{equation}
\begin{equation}\label{vhot}
    v_{\rm{wind,\,h}} \approx 0.2v_{\rm{K}}(r_{\rm tr}),
\end{equation}
where $\dot{M}_{\rm{inflow}}(r_{\rm tr})$ is the accretion rate at the transition radius $r_{\rm tr}$, which is roughly equal to $\dot{M}_{\rm{Bondi}}$. $v_{\rm{K}}(r_{\rm tr})$ is the Keplerian velocity at the transition radius. 

Combining equations~\eqref{rtr}, \eqref{mdoth} and \eqref{vhot}, and using 
$\dot{E}_{\rm wind,h}=\frac{1}{2}\dot{M}_{\rm wind,\ h}v_{\rm wind,\ h}^2$, 
the wind energy output from the hot accretion flow can be written as
\begin{equation}\label{hote}
    \dot{E}_{\rm{wind,\,h}}\approx 8.3\, \dot{m}_{\rm{Bondi}}^2(1-50\,\dot{m}_{\rm{Bondi}})\,\dot{M}_{\rm{Bondi}}c^2,
\end{equation}
where $\dot{m}_{\rm{Bondi}}\equiv\dot{M}_{\rm{Bondi}}/\dot{M}_{\rm{Edd}}$ is the Eddington ratio of the accretion rate. In the hot accretion flow, a large part of the accreting material will finally be converted to a wind and flow out, so the accretion fraction should be treated carefully. Combining Eqs.~(\ref{rtr}) and (\ref{mdoth}), and following \citet{yuan18}, the accretion fraction in the hot mode can be described as
\begin{equation}\label{facc}
    f_{\rm{acc}}=\frac{\dot{M}_{\rm{BH}}}{\dot{M}_{\rm{Bondi}}}\approx(3r_{\rm{s}}/r_{\rm{tr}})^{0.5}\approx\frac{\dot{m}_{\rm{Bondi}}}{2\times10^{-2}}.
\end{equation}
The lower right panel of Fig.~\ref{fig:mmodel} shows the corresponding accretion fraction as a function of the accretion rate. This accretion fraction has recently garnered attention in the literature. For example, \citet{su25} incorporated a suppressed fraction in their galaxy-scale simulations to demonstrate its critical role in regulating black hole mass growth and galaxy evolution, particularly when combined with AGN feedback.

\subsubsection{Jet energy in the hot mode}

\begin{figure*}
 \subfigure{\includegraphics[width=0.45\textwidth]{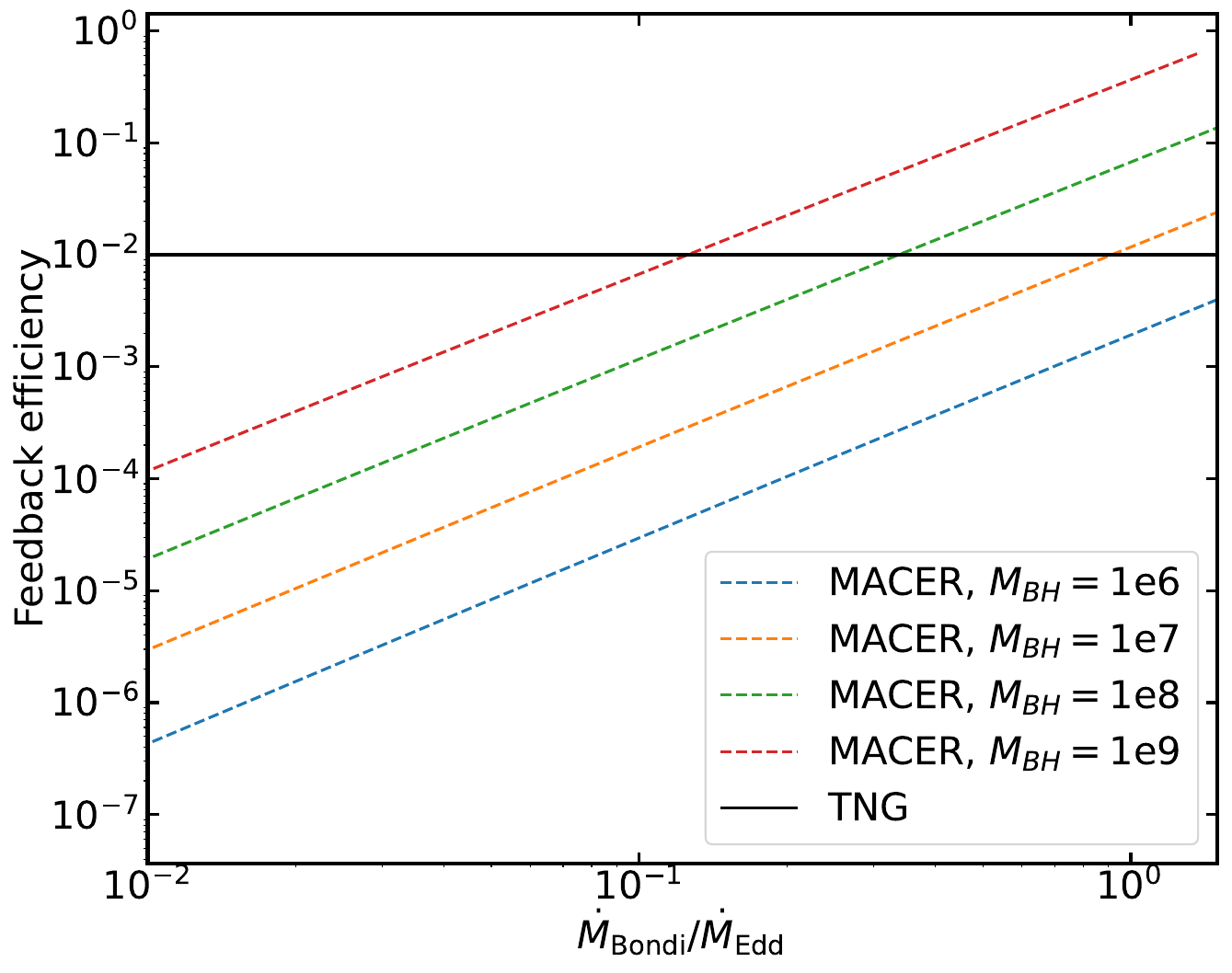}}
 \subfigure{\includegraphics[width=0.45\textwidth]{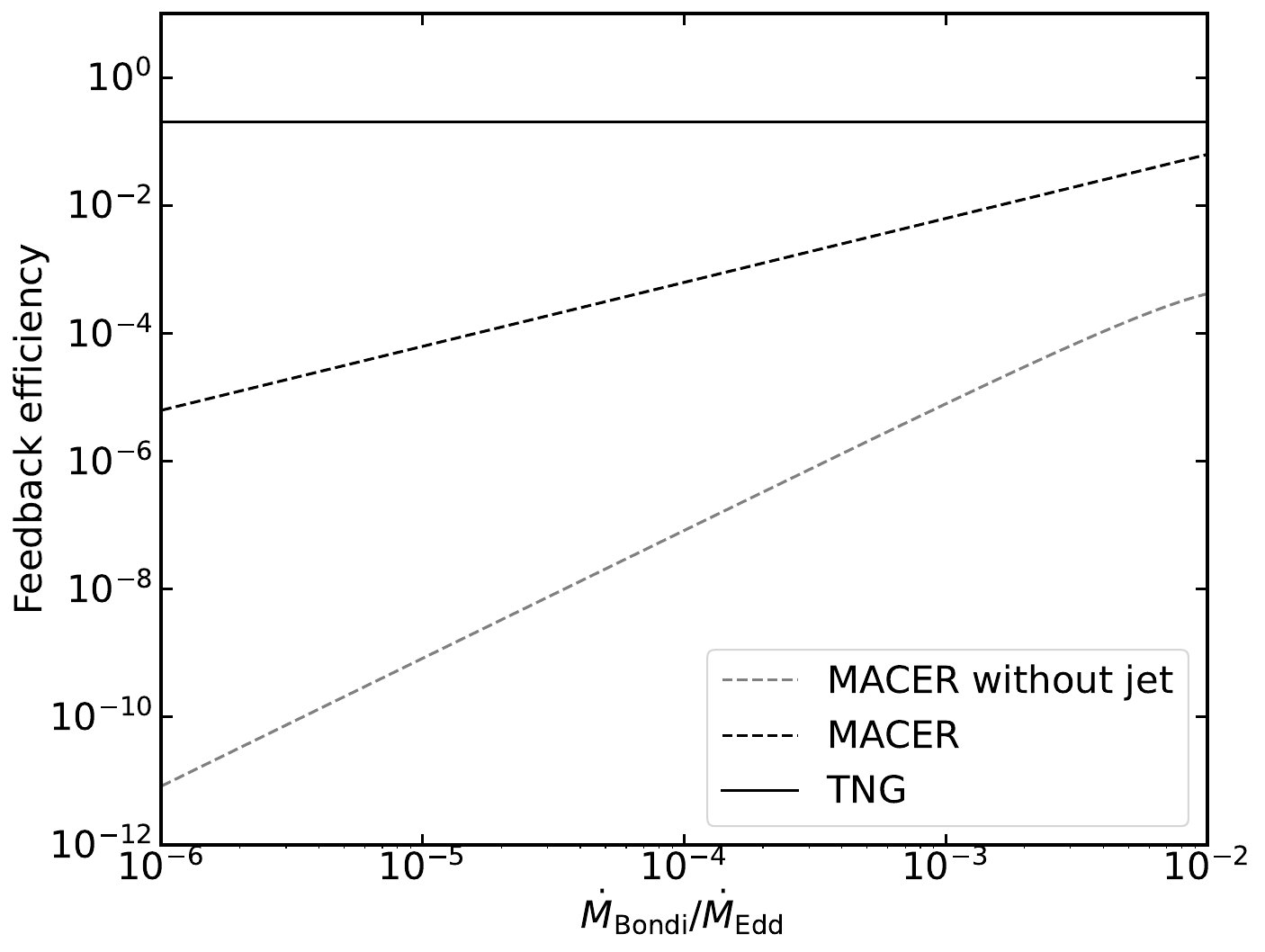}}\\
 \subfigure{\includegraphics[width=0.45\textwidth]{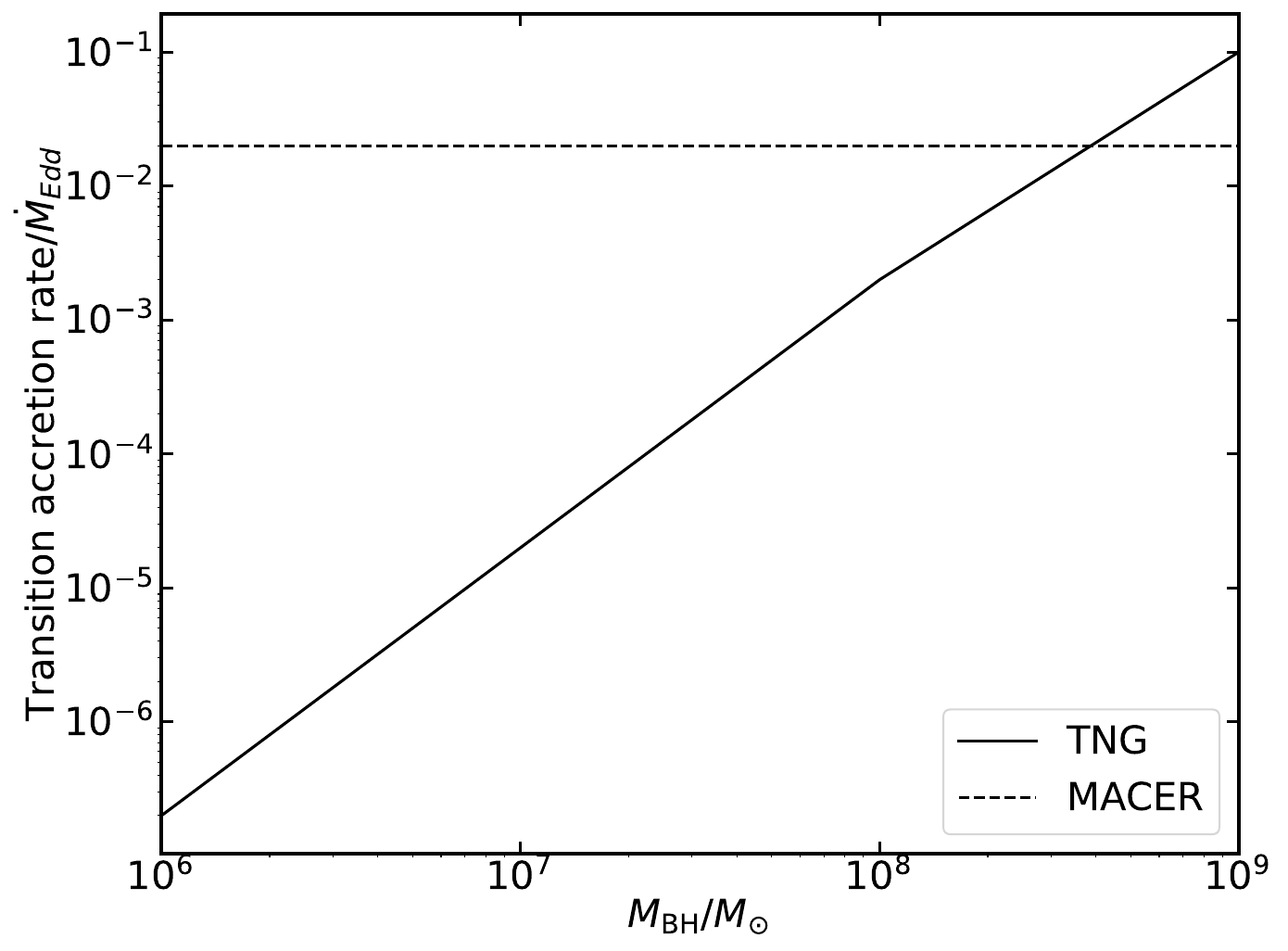}}
 \subfigure{\includegraphics[width=0.45\textwidth]{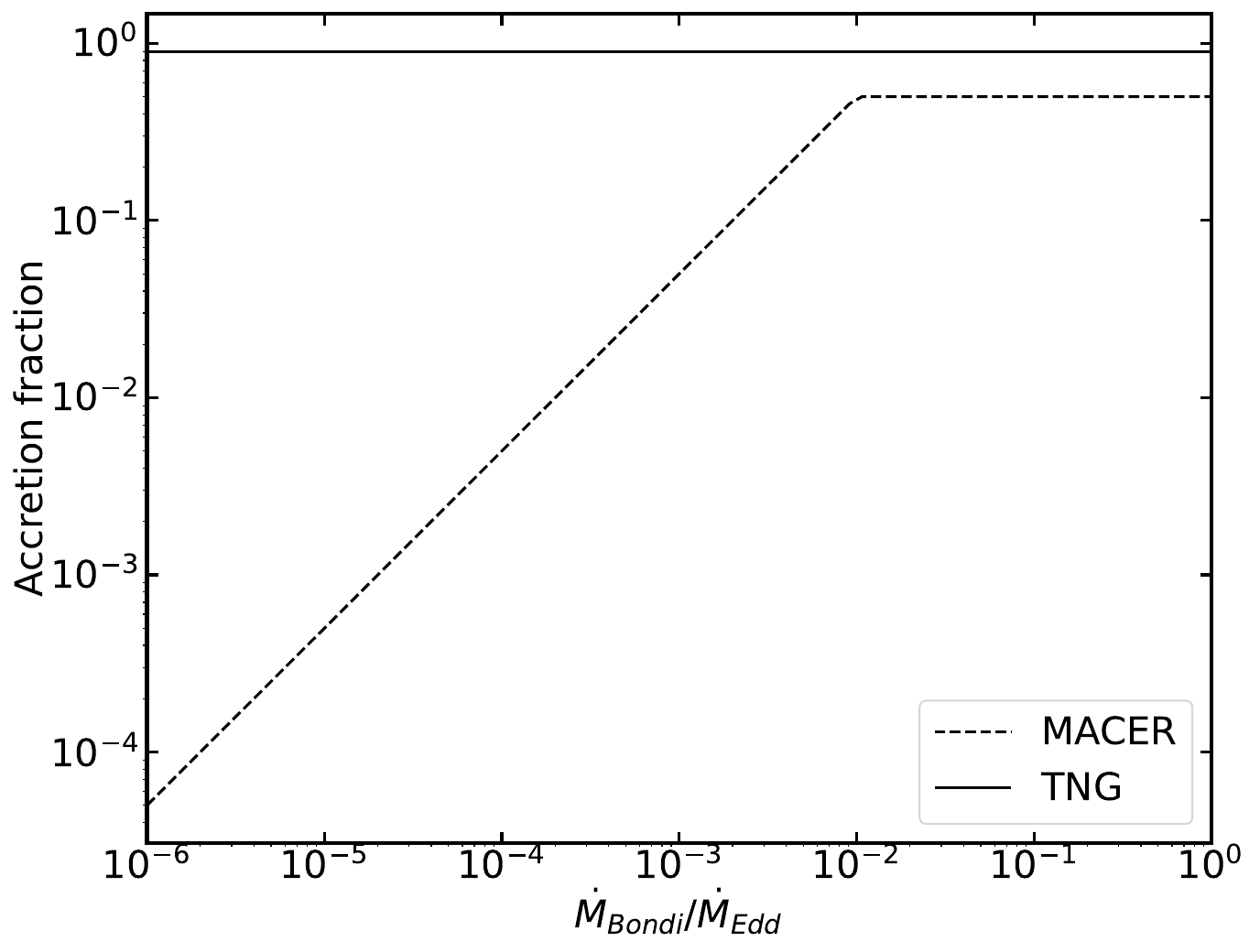}}
 \caption{{\it Upper left:} Feedback efficiency versus Bondi accretion rate in the high–accretion-rate regime (quasar/cold mode) for basic-MACER and TNG. Dotted curves (different colours) show basic-MACER at different $M_{\rm BH}$; the solid curve shows TNG for $M_{\rm BH}\approx2.236\times10^8\,{\rm M_{\odot}}$. The feedback efficiency is defined as $\dot{E}/(\dot M_{\rm Bondi} c^2)$. {\it Upper right:} Feedback efficiency at low accretion rates (radio/hot mode): basic-MACER with and without jets (black/gray dotted), and TNG (solid). The definition of feedback efficiency is the same as in the upper left panel. {\it Lower left:} Transition accretion rate in units of the Eddington rate between feedback modes as a function of $M_{\rm BH}$ in basic-MACER and TNG. {\it Lower right:} Accretion fraction versus accretion rate in basic-MACER and TNG.}
 \label{fig:mmodel}
\end{figure*}

Next to the winds, jets remain an important part of AGN feedback. The observed radio jets in galaxy clusters are believed to suppress cooling flows \citep[see][for a review]{fabian12}. In the original MACER model, jets are not included, but we consider them in the basic-MACER model. However, in cosmological simulations the spatial resolution is normally insufficient to resolve jets directly, so we only consider the jet energy based on theoretical works and inject it along with the wind feedback using the numerical wind injection implementation in TNG. \citet{yang21} performed a detailed analysis of the jet energy from a hot accretion flow based on GRMHD simulations that varied the black hole spin and the magnetization of the hot accretion flow. Referring to \citet{yang21}, the $\dot{M}_{\rm jet}$ is the same order of $\dot{M}_{\rm BH}$. For simplicity, we parameterize the jet mass outflow rate and velocity as
\begin{equation}\label{mdotjet}
    \dot{M}_{\rm{jet}} \simeq \dot{M}_{\rm{BH}} = f_{\rm{acc}}\dot{M}_{\rm{Bondi}}
\end{equation}
with
\begin{equation}\label{vjet}
    v_{\rm{jet}}=0.5\,c
\end{equation}
in the basic-MACER model. Note that the total jet power is smaller than the power obtained in \citet{yang21} for a rapidly spinning black hole with an MAD since we assume the black hole is moderately spinning. Combining equations \eqref{facc}, \eqref{mdotjet}, and \eqref{vjet}, the energy output from jets can then be described as
\begin{equation}\label{hotjete}    \dot{E}_{\rm{jet}}=6.25\,\dot{m}_{\rm{Bondi}}\dot{M}_{\rm{Bondi}}c^2.
\end{equation}
We note that GRMHD simulations demonstrate a strong dependence of the jet efficiency on black hole spin and magnetic flux state \citep[e.g. MAD vs. SANE;][]{tchekhovskoy10,yang21,narayan2022}. The jet efficiency implied by our prescription is $\eta_{\rm jet}\simeq 0.125$. Since the jet power scales roughly as $a_*^2$ at moderate spins, this corresponds to $a_*\simeq0.4$. In the present cosmological framework, neither the black hole spin nor magnetic flux evolution is followed self-consistently. The adopted parameterization should therefore be regarded as an effective prescription representative of typical hot-mode accretion conditions, rather than a detailed model of spin-dependent jet launching. As illustrated in Fig.~\ref{fig:mmodel}, jets carry more energy than winds, but their much narrower opening angle implies a lower coupling efficiency with the interstellar medium (ISM). The limited spatial resolution of our cosmological resolutions does not permit an explicit treatment of the angular dependence, which should be explored in future high-resolution studies.

In this work, the basic-MACER model includes the AGN wind, jet and radiation feedback. The numerical implementation of kinetic energy feedback (e.g., wind and jet) follows the TNG prescription, which is described in Sec.~\ref{sec:kfbi}. The radiation feedback from AGN follows the Illustris and TNG models, with a detailed description given in \citet{vogelsberger14}. The radiation feedback from AGN is assumed to be a point radiation source with a \citet{korista1997} AGN spectral energy distribution. Some fraction of the radiation will be obscured when it leaves the central region, where the ``observable fraction'' of the AGN luminosity is $f_{\rm bol}^{\rm AGN,\,obs}=\omega_1(L_{\rm bol}^{\rm AGN}/10^{46}{\rm erg\,s^{-1}})^{\omega_2}$, with $\omega_1=0.3$ and $\omega_2=0.07$. The gas in the galaxy is assumed to be optically thin for the AGN radiation except for the star-forming gas.

\subsection{The AGN feedback model in TNG}\label{TNG}

The AGN feedback model in TNG is described comprehensively by \citet{weinberger17}; we here only briefly summarize its main components.

Similar to MACER, TNG adopts a two-mode feedback scenario. The two feedback modes switch at a critical Eddington ratio threshold $\chi$ that is scaled by the black hole mass,
\begin{equation}\label{trans}
\chi = \min\left[\chi_0\left(\frac{M_{\rm BH}}{10^8\,{\rm M}_\odot}\right)^{\beta},\,0.1\right],
\end{equation}
where $\chi_0=0.002$ and $\beta=2$.
In the quasar mode (for high accretion rate), an AGN deposits thermal energy into the surrounding gas, with a rate
\begin{equation}\label{tnghigh}
\dot{E}_{\rm high} = \epsilon_{\rm f,high}\,\epsilon_r\,\dot{M}_{\rm Bondi}c^2,
\end{equation}
with $\epsilon_r=0.2$ and $\epsilon_{\rm f,high}=0.1$. The energy is distributed to neighboring cells using the same kernel weighting as in the accretion estimate.

In the radio mode (i.e.~for low accretion rate), the AGN injects kinetic energy at a rate
\begin{equation}
\dot{E}_{\rm low} = \epsilon_{\rm f,kin}\,\dot{M}_{\rm Bondi}c^2,
\end{equation}
where the efficiency 
\begin{equation}\label{TNGLow}
\epsilon_{\rm f,kin} = \min\left(\frac{\rho}{f_{\rm thresh}\,\rho_{\rm SFthresh}},\,0.2\right)
\end{equation}
depends on the local gas density,
where $f_{\rm thresh}=0.05$, and $\rho_{\rm SFthresh}$  designate the star formation density threshold.

\subsection{Comparison of two AGN feedback models}\label{comparison}

It is instructive to briefly recall the main differences between the MACER and TNG models for black hole feedback physics. In the original MACER project, the outer boundary of the black hole accretion flow is explicitly resolved, i.e.~the inner boundary of the simulation lies inside the accretion-flow's outer boundary (near the Bondi radius). This allows the mass flux through the inner boundary to be measured and, combined with accretion theory, to be translated into the horizon-scale accretion rate. Implementing this approach in TNG is computationally prohibitive at present (although it may be attempted in the future with localized refinement around the BH), so in the present work we retain TNG’s original, empirical accretion-rate estimate. Beyond the accretion-rate estimation, MACER and TNG differ in several key aspects as summarized below.

\begin{itemize}
\item \textbf{Feedback channel and energy deposition at high accretion rates.}
In basic-MACER, quasar-mode feedback is dominated by a kinetic wind whose interaction with the ISM is modeled self-consistently; the deposited energy emerges from the wind dynamics. In TNG, quasar-mode feedback is empirical: a fixed fraction of the AGN power is assumed to couple thermally to the BH surroundings.

\item \textbf{Feedback efficiency at high accretion rates.}  
In TNG, the power of the quasar-mode follows Eq.~\eqref{tnghigh} with a constant efficiency of $\epsilon=\dot{E}/(\dot{M}c^2)$, independent of $M_{\rm BH}$ and $\dot{M}_{\rm BH}$. In basic-MACER, the wind power follows Eqs.~\eqref{coldwindflux}–\eqref{colde}, calibrated to UFO observations, so the effective efficiency varies with $M_{\rm BH}$ and accretion rate: it decreases toward lower $\dot{M}_{\rm Bondi}$ and smaller $M_{\rm BH}$  since the low-mass BH have low luminosity at fixed $\dot{M}_{\rm Bondi}/\dot{M}_{\rm Edd}$. The upper-left panel of Fig.~\ref{fig:mmodel} compares these scalings, showing that for $M_{\rm BH}\lesssim10^7\,{\rm M}_\odot$ the efficiency of basic-MACER lies generally below TNG’s constant value. Because strong kinetic feedback at high accretion rates can curtail BH growth, these differences directly impact the $M_\star$–$M_{\rm BH}$ relation.

\item \textbf{Feedback efficiency at low accretion rates.} 
Both models inject kinetic energy in the radio/hot mode, but the efficiencies differ. TNG adopts a density-dependent efficiency [Eq.~\eqref{TNGLow}] with an upper cap of 0.2, implicitly assuming full coupling to the local ISM up to that limit. In basic-MACER, kinetic feedback combines winds [Eq.~\eqref{hote}] and jets [Eq.~\eqref{hotjete}], yielding an efficiency that varies with the Bondi Eddington ratio. As shown in the upper-right panel of Fig.~\ref{fig:mmodel}, the total efficiency of basic-MACER is systematically below TNG’s over most of the low-$\dot{M}_{\rm Bondi}$ range.

\item \textbf{Mode–transition criterion.}  
Both models are two-mode, but the transition criterion differs significantly. TNG uses a BH-mass–dependent threshold $\chi$ [Eq.~\eqref{trans}], capped at 0.1, whereas basic-MACER adopts a fixed transition at $L/L_{\rm Edd}\simeq0.01$ (i.e., of order $\sim1\%$ in $\dot{M}_{\rm Bondi}/\dot{M}_{\rm Edd}$) motivated by X-ray binaries and accretion theory. The lower-left panel of Fig.~\ref{fig:mmodel} shows that in TNG the transition reaches $\sim1\%$ only for $M_{\rm BH}\gtrsim10^8\,{\rm M}_\odot$, effectively delaying the entry into the kinetic mode for low-mass BHs, thereby facilitating their early growth while favoring a transition to the kinetic mode for massive BHs, where they then grow more slowly.

\item \textbf{Accretion fraction onto the BH.}  
Because winds/jets remove mass, not all large-scale inflow reaches the BH. In basic-MACER, the accretion fraction is $0.5$ in the cold mode, and follows Eq.~\eqref{facc} in the hot mode, decreasing sharply at low $\dot{m}_{\rm Bondi}$. In TNG, a fixed fraction equal to $0.9$ is assumed instead (i.e., $10\%$ of the rest mass is converted into feedback). The lower-right panel of Fig.~\ref{fig:mmodel} shows that the accretion fraction of basic-MACER rises with $\dot{M}$ but remains below TNG’s across most of the parameter space. In fact, in the hot mode it can be extremely low, implying very limited BH growth at low accretion rates.
   
\end{itemize}

\begin{table*}
 \caption{Summary of model names and their corresponding model characteristics. The `basic-MACER' and `TNG' models refer to our default versions that we primarily compare, whereas the other models are variations that we use to investigate parameter dependencies.}
 \begin{tabular*}{\textwidth}{@{}l@{\hspace*{5pt}}|l@{\hspace*{20pt}}l@{\hspace*{20pt}}l@{\hspace*{20pt}}l@{}}
 \hline
 Model name & \makecell[c]{quasar mode} & \makecell[c]{radio mode} & \makecell[c]{transition accretion rate} \\
 \hline
 basic-MACER & \makecell[c]{kinetic injection;\\efficiency changing with BHAR and $M_{\rm BH}$;\\constant accretion fraction} & \makecell[c]{kinetic injection;\\efficiency changing with BHAR;\\accretion fraction changing with BHAR} & \makecell[c]{constant value} \\
  \hline
 TNG & \makecell[c]{thermal injection;\\ constant efficiency;\\constant accretion fraction} & \makecell[c]{kinetic injection;\\constant efficiency;\\constant accretion fraction} & \makecell[c]{changing with $M_{\rm BH}$}\\
  \hline
 ThermalHigh& \makecell[c]{the same as TNG} & \makecell[c]{the same as basic-MACER} & \makecell[c]{the same as basic-MACER} \\
  \hline
 FixHigh& \makecell[c]{the same as basic-MACER \\ except for fixing the efficiency at $M_{\rm BH}=10^8~{\rm M_{\odot}}$} & \makecell[c]{the same as basic-MACER} & \makecell[c]{the same as basic-MACER}\\
  \hline
 OnlyWind& \makecell[c]{the same as basic-MACER} & \makecell[c]{ignore the contribution from jet} & \makecell[c]{the same as basic-MACER} \\
  \hline
 HighLowEff& \makecell[c]{the same as basic-MACER} & \makecell[c]{the same as TNG} & \makecell[c]{the same as \\ basic-MACER} \\
  \hline
 HighAcc& \makecell[c]{the same as basic-MACER\\ except for with accretion fraction equal to 0.9} & \makecell[c]{the same as basic-MACER\\ except that the accretion fraction is equal to 0.9} & \makecell[c]{the same as basic-MACER} \\
 
 \hline
 \end{tabular*}
 \label{table:des}
\end{table*}

\begin{table*}
 \caption{Summary of the cosmological simulation runs performed in this work. For each simulation, we list the run name, the adopted feedback model (see Table 1 for physical details), the box size, the number of particles ($N_{\rm particle}$), the gravitational softening lengths for dark matter and gas ($\epsilon_{\rm DM}$ and $\epsilon_{\rm Gas}$), and their mass resolutions ($m_{\rm DM}$ and $m_{\rm Gas}$).}
 \begin{threeparttable}
 \begin{tabular*}{\textwidth}{l|ccccccccc}
  \hline
  Simulation name & \makecell[c]{feedback model} &\makecell[c]{Box size [${\rm Mpc}/h$]} & \makecell[c]{$N_{\rm particle}$} & \makecell[c]{$\epsilon_{\rm DM}$ [$c{\rm kpc}/h$]} & \makecell[c]{$\epsilon_{\rm Gas}$ [$c{\rm kpc}/h$]} & \makecell[c]{$m_{\rm DM}$ [${\rm M_{\odot}}$]} & \makecell[c]{$m_{\rm Gas}$ [${\rm M_{\odot}}$]} \\
   \hline
  basic-MACER& basic-MACER &  50  & 2 $\times 512^3$ & 2& 2 & $6.27\times10^7$ & $1.26\times10^7$\\

  TNG& TNG & 50 & 2 $\times 512^3$ & 2& 2 & $6.27\times10^7$ & $1.26\times10^7$\\

  basic-MACER\_HighRes & basic-MACER &  25  & 2 $\times 512^3$ & 1 & 1 & $7.84\times10^6$ & $1.58\times10^6$\\

  TNG\_HighRes & TNG & 25 & 2 $\times 512^3$ & 1 & 1 & $7.84\times10^6$ & $1.58\times10^6$\\

  ${\rm basic-MACER}^*$ & basic-MACER &  25  & 2 $\times 256^3$ & 2& 2 & $6.27\times10^7$ & $1.26\times10^7$\\

  ${\rm TNG}^*$ & TNG & 25 & 2 $\times 256^3$ & 2& 2 & $6.27\times10^7$ & $1.26\times10^7$\\

  ThermalHigh & ThermalHigh &  25 & 2 $\times 256^3$& 2& 2 & $6.27\times10^7$ & $1.26\times10^7$\\

  FixHigh & FixHigh & 25 & 2$\times 256^3$& 2& 2 & $6.27\times10^7$ & $1.26\times10^7$\\

  OnlyWind & OnlyWind & 25 & 2 $\times 256^3$& 2& 2 & $6.27\times10^7$ & $1.26\times10^7$\\

  HighLowEff & HighLowEff & 25 & 2 $\times 256^3$& 2& 2 & $6.27\times10^7$ & $1.26\times10^7$\\

  HighAcc & HighAcc & 25 & 2 $\times 256^3$& 2& 2 & $6.27\times10^7$ & $1.26\times10^7$\\
   \hline 
 \end{tabular*}
 \label{table:run}
 \begin{tablenotes}
      \footnotesize 
      \item[*] These denote smaller-box simulations used exclusively in the parameter study (Section \ref{sec:param}). Their initial conditions are aligned with other test simulations to ensure consistent comparisons. 
    \end{tablenotes}
 \end{threeparttable}
 \end{table*}

\section{Simulation setup}\label{sec:setup}

AGN feedback is considered to be a key process in the evolution of massive systems such as elliptical galaxies. To investigate the impact of  different feedback models considered here, we perform both idealized simulations of an isolated elliptical galaxy and cosmological box simulations. This section describes the setup of both the elliptical galaxy and the cosmological runs.

\subsection{Idealized elliptical galaxy simulations}

We construct a template system with a halo mass of $10^{13}\,h^{-1}{\rm M_{\odot}}$, representing a massive elliptical galaxy. The dark matter halo follows a Navarro–Frenk–White (NFW) profile \citep{nfw96},
\begin{equation}
\rho(r) = \frac{\rho_0}{(r/r_s)(1+r/r_s)^2},
\end{equation}
where $\rho_0$ is determined such that the integrated halo mass within the virial radius equals $10^{13}\,h^{-1}{\rm M}_\odot$. The corresponding virial radius is $\sim350.4\,h^{-1}{\rm kpc}$. The scale radius is defined as $r_s \equiv R_{\rm vir}/c$, where the concentration parameter is set to $c=8.08$, following the fitting relation of \citet{zhao09}.

The stellar component is modeled with a Hernquist profile \citep{hernquist90},
\begin{equation}
\rho_\star(r) = \frac{M_\star}{2\pi r_\star^3}\,\frac{1}{(r/r_\star)(1+r/r_\star)^3},
\end{equation}
with a total stellar mass $M_\star = 10^{11}\,h^{-1}{\rm M_\odot}$ and scale length $r_\star$ corresponding to an effective radius $r_{\rm eff} = 4.17\,h^{-1}{\rm kpc}$ ($r_{\rm eff}=1.8153\,r_\star$). A central supermassive black hole of mass $4.9\times10^8\,{\rm M_\odot}$ is placed at the halo center. The black hole mass is set to follow the scaling relation reported in \citet{kormendy13}. The combined gravitational potential of the stars and the dark matter is treated as static. The initial gas distribution follows a $\beta$-model profile,
\begin{equation}
\rho_{\rm gas}(r) = \rho_0 \left[1+(r/r_0)^2\right]^{-1.5\,\beta},
\end{equation}
with $\rho_0 = 0.4\,{\rm cm^{-3}}$, $\beta=0.5$, and $r_0=1\,{\rm kpc}$, resembling the observed gas distribution of NGC~5044 \citep{werner14}. The gas angular momentum is prescribed to follow a cylindrical distribution aligned with the $z$-axis. The rotation profile is adopted from \citet{bullock01},
\begin{equation}
v_{\rm rot}(R) = \frac{j_0}{R}\,\frac{m}{\mu-m},
\end{equation}
where $\mu=1.5$, $m=M(<R)/M_{\rm vir}$, $R$ is the cylindrical radius, $M(<R)$ is the enclosed mass within the cylindrical radius, $M_{\rm vir}$ is the virial mass, and $j_0 = \sqrt{2}\,\lambda R_{\rm vir}V_{\rm vir}/[-\mu\log(1-\mu^{-1})-1] = 1.07\times10^4\,{\rm kpc\,km\,s^{-1}}$, and is chosen such that the spin parameter matches $\lambda=0.04$, consistent with tidal torque theory \citep{peebles1969}. 
The metallicity of the gas is set to 0. We note that real elliptical galaxies are metal-enriched \citep{2023MNRAS.524.5787Z}. However, since at the typical temperatures of hot halos ($T>10^6\,{\rm K}$) cooling is dominated by bremsstrahlung, the impact of metal-line cooling is subdominant for the global gas dynamics. The temperature profile is set by assuming hydrostatic equilibrium.
The mass resolution is $10^6\,h^{-1}{\rm M_\odot}$, comparable to that of the TNG100-1 simulation.

\subsection{Cosmological simulations}\label{cosset}

We perform two cosmological simulations in a periodic box of size $50\,h^{-1}{\rm Mpc}$ with different feedback models as our fiducial simulations. Although this volume does not contain many halos with $M_{\rm DM} > 10^{14}\,{\rm M_\odot}$, it provides a sufficient statistical sample in the range $10^{12.5}$–$10^{13}\,{\rm M_\odot}$  to investigate the performance of the different AGN feedback models, which are becoming particularly important at these mass scales.

The adopted cosmology follows the \textit{Planck} intermediate results \citep{planck16}, consistent with TNG: $\Omega_\Lambda=0.6911$, $\Omega_m=0.3089$, $h=0.6774$, $\Omega_b=0.0486$, $n_s=0.9667$, and $\sigma_8=0.8159$. Initial conditions are generated using the \citet{eisenstein98} transfer function. The default resolution employs $512^3$ dark matter particles and an equal number of gas cells in a $50\,{\rm Mpc}/h$ periodic box, corresponding to mass resolutions of $4.25\times10^7\,h^{-1}{\rm M_\odot}$ (dark matter) and $8.38\times10^6\,h^{-1}{\rm M_\odot}$ (gas). This is comparable to the TNG100-2 simulation. The gravitational softening length is 2 comoving kpc (capped at 1 proper kpc) for stars and dark matter, while for gas cells it is adaptive with a minimum of 0.25 comoving kpc. At this resolution, galaxy properties are reliably captured for halos above $M_{\rm halo}\sim10^{12.5}\,{\rm M_\odot}$, although convergence is not guaranteed for lower-mass systems. 

To further isolate the effects of different parameter choices for the AGN feedback models, we also run a suite of smaller test simulations. These use $2\times256^3$ particles in a $25\,h^{-1}{\rm Mpc}$ box, sharing the same initial conditions. For comparison, we also run the simulations with the basic-MACER and TNG models using the same initial conditions; these two models are only used in Sec.~\ref{sec:param}. These test model and the simulation setup are summarized in Table~\ref{table:des} and Table~\ref{table:run}. All simulations were performed with the TreePM magneto-hydrodynamical moving-mesh code {\small AREPO} \citep{springel10}.

\section{Results from idealized elliptical galaxies}\label{sec:idealized}

In this section, we examine the performance of the basic-MACER and TNG models in our idealized elliptical galaxy simulations. For comparison, we also include a corresponding simulation without AGN feedback.

\subsection{Star formation rate and black hole accretion rate}

The SFR and BHAR are two key diagnostics for assessing the effectiveness of an AGN feedback model in elliptical galaxies. In this section, we therefore consider the behaviour of these quantities in our idealized elliptical galaxy simulations when the  basic-MACER and TNG models are applied. 

\begin{figure}
 \subfigure{\includegraphics[width=\columnwidth]{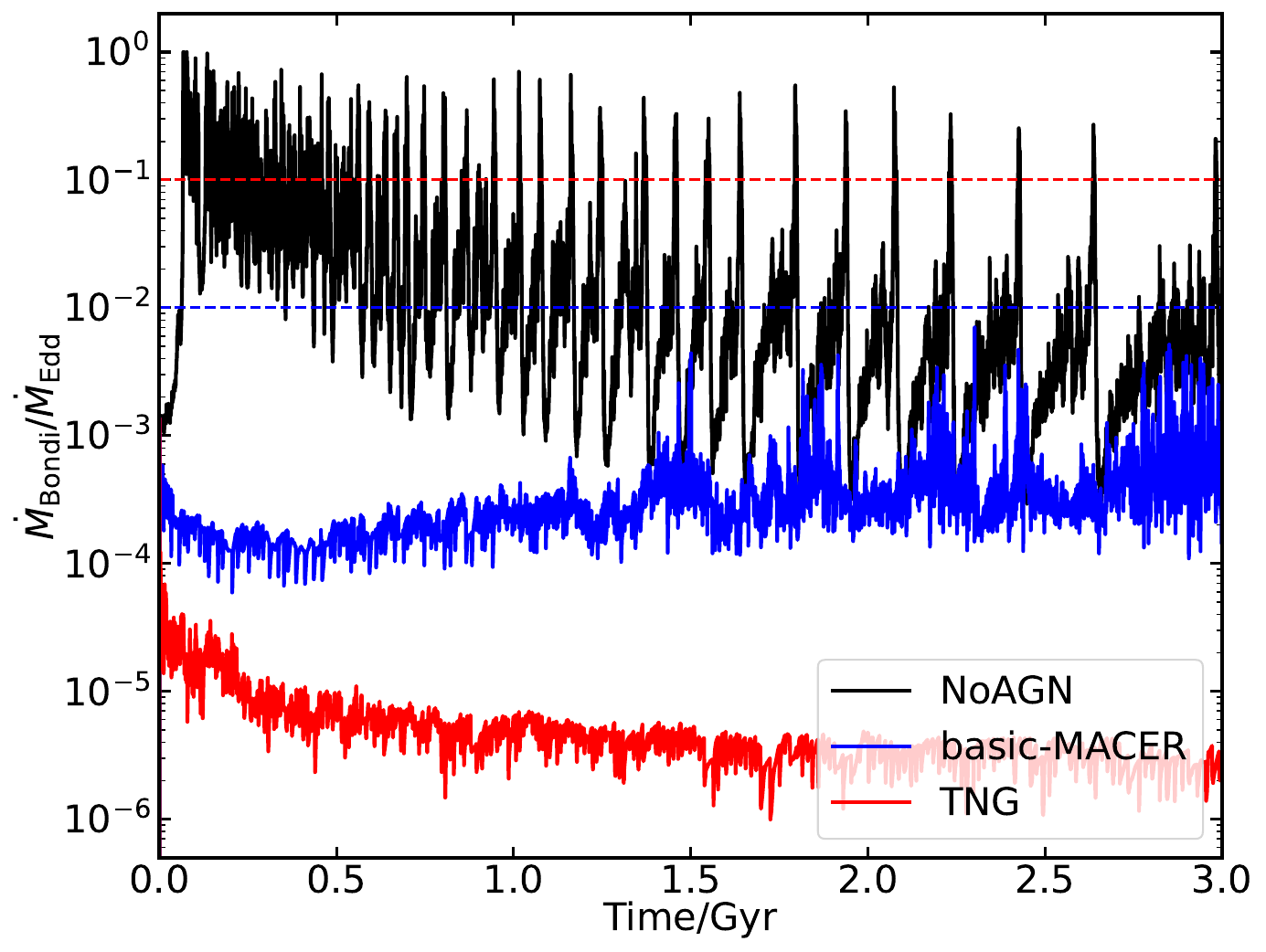}}\\
 \subfigure{\includegraphics[width=\columnwidth]{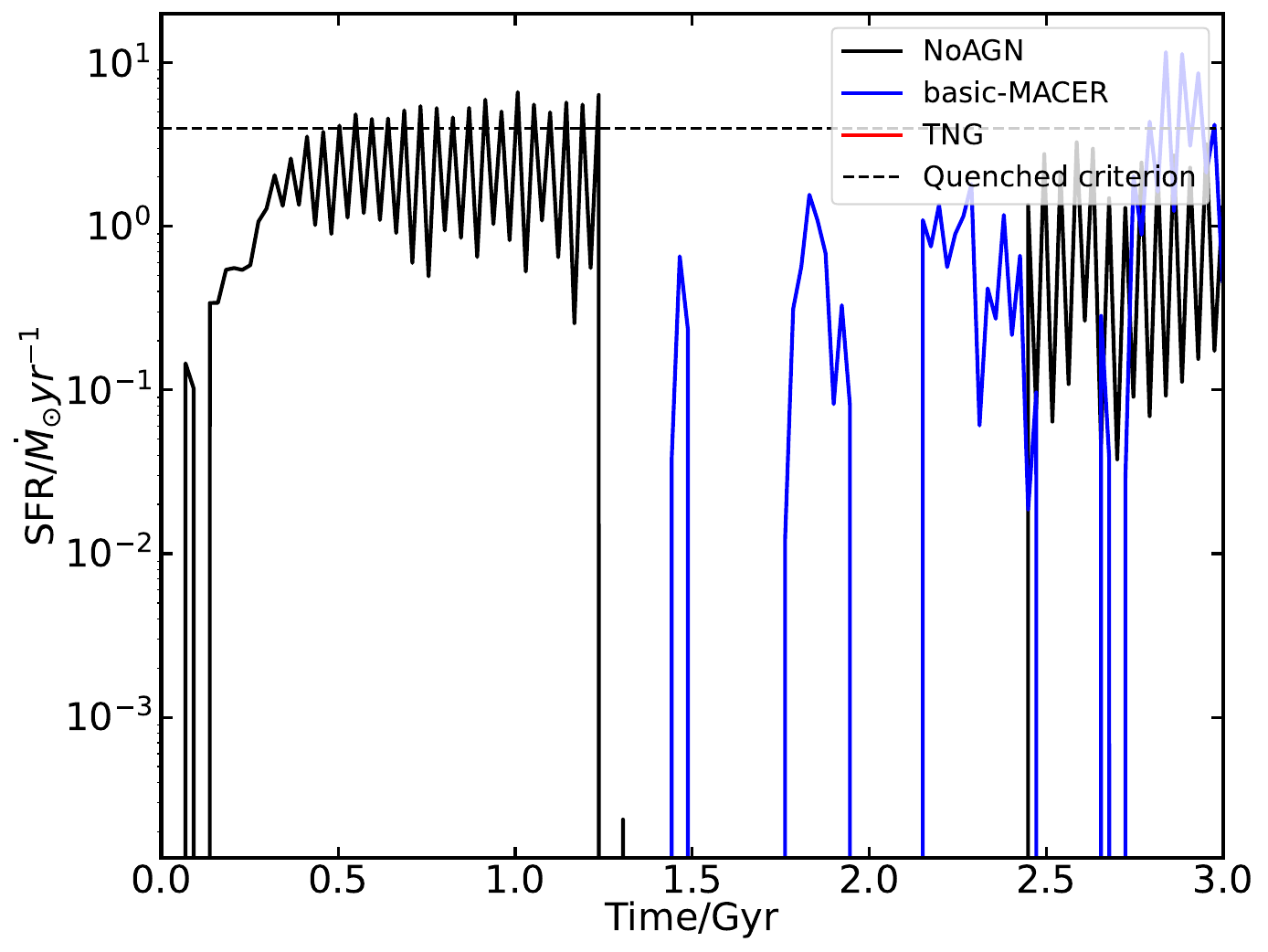}}\\
 \subfigure{\includegraphics[width=\columnwidth]{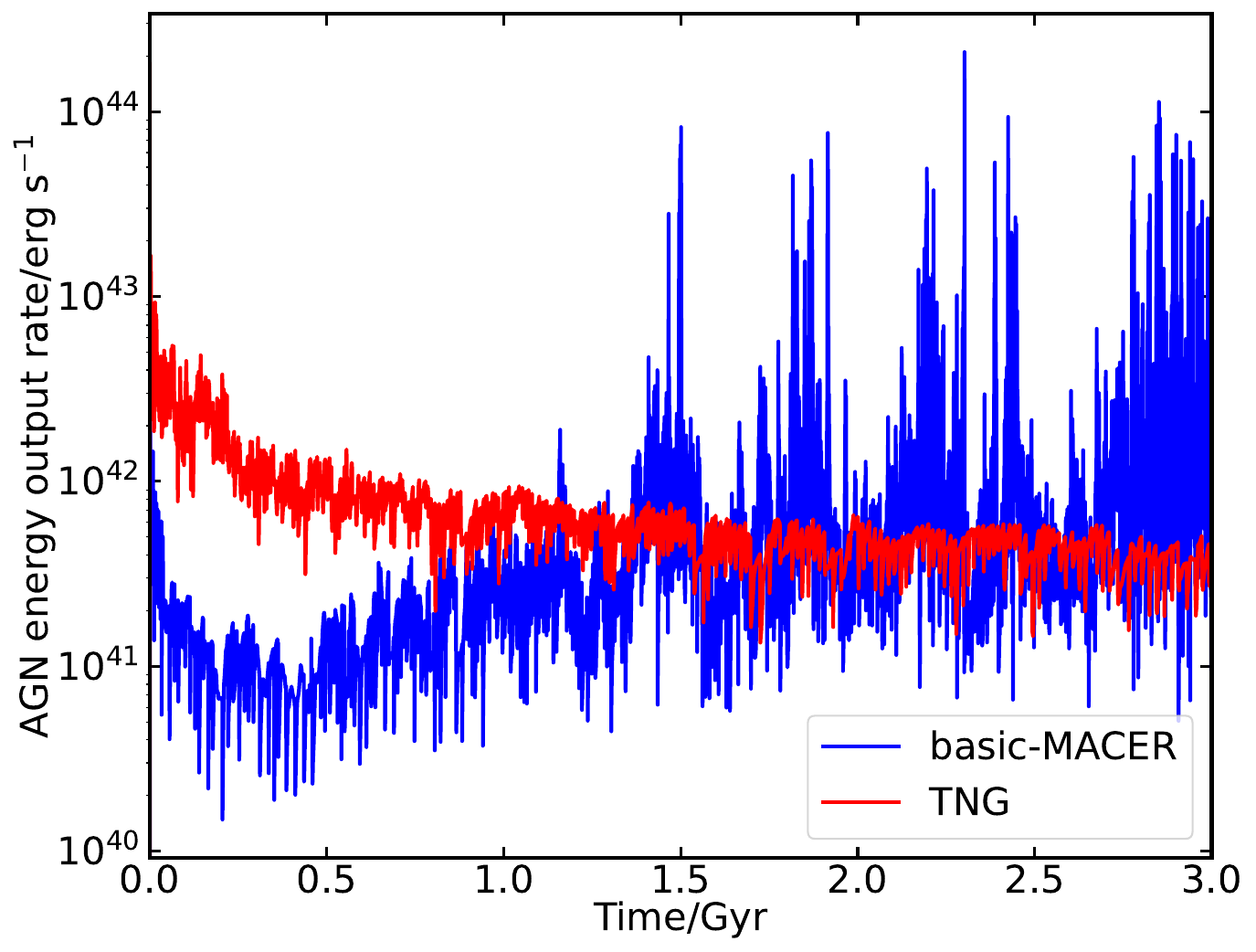}}
 \caption{Bondi accretion rate ({\it upper panel}), SFR ({\it middle panel}) and AGN energy output rate ({\it lower panel}) as a function of simulation time for three models (without AGN feedback, basic-MACER, and TNG) in the idealized elliptical galaxy simulation. The dashed lines in the upper panel represent the transition accretion rate for the two models. The dashed line in the lower panel represents the quenching criterion for star formation. In the simulated galaxy with the TNG model, the star formation rate is identical to zero. These figures show that the simulated elliptical galaxy with the TNG model has a substantially lower Bondi accretion rate and SFR than with the basic-MACER model.}
 \label{fig:imdot&sfr}
\end{figure}

The upper panel of Fig.~\ref{fig:imdot&sfr} shows the Bondi accretion rate as a function of time, with the dashed line representing the transition accretion rate for both models. The figure shows that black holes in both models stay at low accretion rates, indicating that the elliptical galaxies in both models are dominated by the low accretion rate mode, consistent with the canonical MACER simulation of elliptical galaxies \citep{yuan18} and with observations \citep{fabian12}. The Bondi accretion rate in basic-MACER is higher than in TNG at all times. This result can be explained by the fact that the released AGN feedback energy should balance the radiative cooling of the hot gaseous halo, as  already discussed in many previous works \citep[e.g., ][]{li15, wang19}. Since the feedback efficiency in basic-MACER is lower than in TNG at the same Bondi accretion rate, the feedback process will adjust the accretion rate and make the energy output rate roughly balance the radiative cooling. So the Bondi accretion rate in basic-MACER ends up being higher than in TNG. 

We note that the NoAGN case shows a gradual decline in $\dot{M}_{\rm Bondi}/\dot{M}_{\rm Edd}$ over time despite maintaining a high Bondi accretion rate. This is primarily due to the continuous growth of the black hole mass in the absence of AGN feedback. Without a mechanism to regulate accretion, the black hole mass increases by nearly two orders of magnitude over 3 Gyr, substantially raising $\dot{M}_{\rm Edd}\propto M_{\rm BH}$ and thus driving down the ratio $\dot{M}_{\rm Bondi}/\dot{M}_{\rm Edd}$. This contrasts sharply with the basic-MACER and TNG cases, where AGN feedback self-regulates the accretion process and keeps the black hole mass nearly constant.

The middle panel of Fig.~\ref{fig:imdot&sfr} shows the SFR as a function of simulation time in both models, and the black dashed line represents a commonly adopted criterion for star formation quenching. In general, both models can produce a quenched galaxy, and in fact, the SFR in both models stays mainly below the quenching criterion. However, the detailed behavior of the SFR in the two models is quite different. In basic-MACER, the SFR exhibits repeated bursts, implying that the elliptical galaxy with the basic-MACER model shows periodic star formation activity, similar to the result in the original MACER. In contrast, in TNG effectively no star formation occurs during the simulation time, and the galaxy stays fully quenched all the time. This difference indicates that the TNG model is more efficient than basic-MACER in  quenching galaxies. 

Previous works \citep[e.g.,][]{prasad15,li15,voit21} have shown that repeated star formation bursts can arise due to the growth of local thermal instabilities and can be explained in terms of an AGN feedback cycle in massive halos. First, the hot gas  cool downs due to radiative cooling. Thermal instability will be triggered when the hot gas meets the thermal instability criterion, i.e., the cooling timescale $t_{\rm{cool}}$ becomes less than roughly $10$ times the free fall timescale $t_{\rm{ff}}$. Part of this cold gas will trigger star formation while the rest will fall into the centre where it can trigger AGN. The entropy of the interstellar medium will then increase due to the feedback heating by the AGN, stifling further growths of the thermal instability. Finally, the gaseous halo returns to being thermally stable. 

In basic-MACER, the repeated star formation bursts can also be explained by the growth of local thermal instabilities. However, we do not find corresponding bursts of the accretion rate. This result may be due to the approach for estimating the AGN accretion rate, which is inherited from TNG in our simulations, and the low resolution of the simulations. Since we calculate the accretion rate by averaging the properties of the gas surrounding the black hole, this treatment will tend to smooth out accretion rate fluctuations. Furthermore, at low resolution  the simulation will not be able to resolve the cold gas well. 

Different from basic-MACER, no star formation appears in TNG. Since star formation needs a supply of cold gas, and such cold gas will arise from the growth of thermal instabilities, this result indicates that the hot gaseous halo never meets the thermal instability criterion due to efficient AGN heating. We will discuss this result further in Section~\ref{radprofile}.

\subsection{Radial profiles of some thermodynamic quantities of the circumgalactic medium}\label{radprofile}

Radial profiles of the thermodynamic quantities of the gas are another critical diagnostic for the impact of AGN feedback. In this section, we therefore analyze the radial profiles of the circumgalactic medium obtained for the  two models. 

\begin{figure*}
\subfigure{\includegraphics[width=1\linewidth]{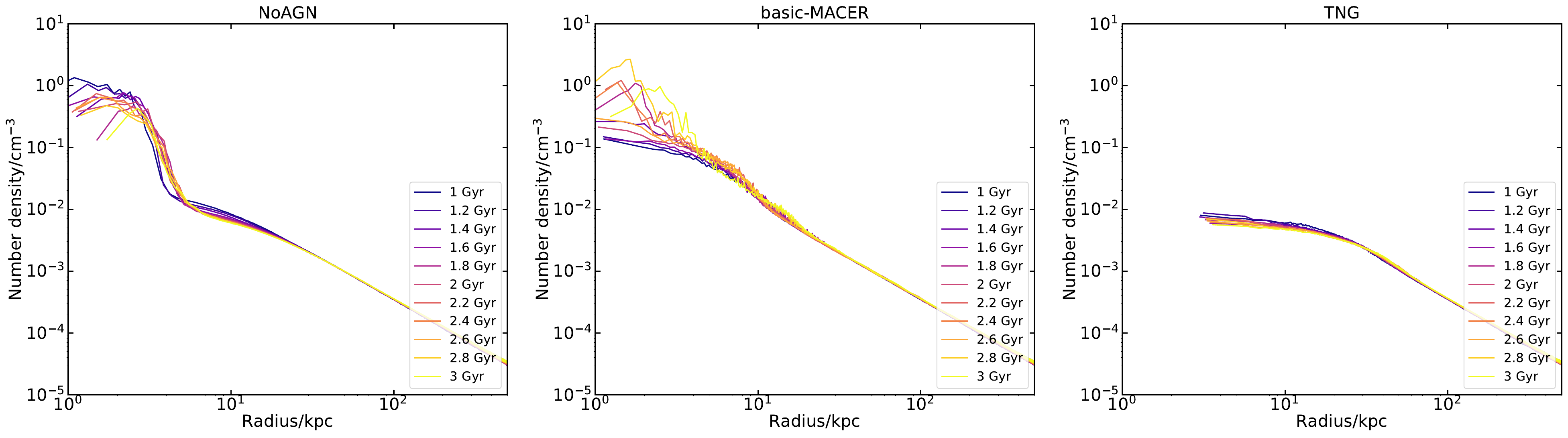}}
\\
\subfigure{\includegraphics[width=1\linewidth]{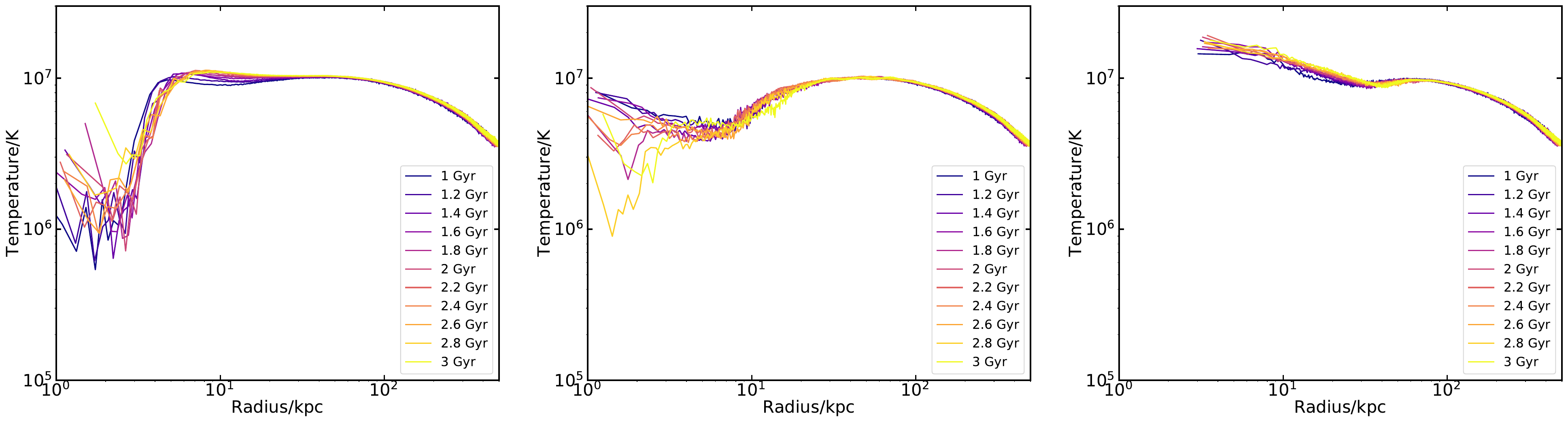}}
\\
\subfigure{\includegraphics[width=1\linewidth]{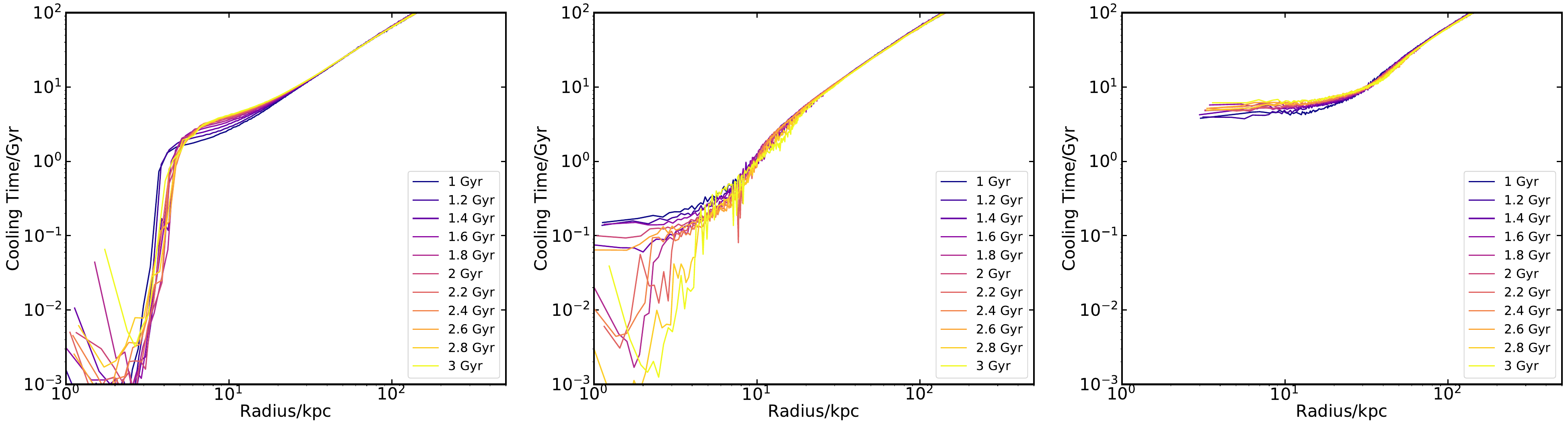}}
\\
\caption{Radial profiles of density, temperature and cooling time for the simulated elliptical galaxy in NoAGN, basic-MACER and TNG. The colour of the lines represents the time at which the radial density profile was calculated. The figure shows that the simulated galaxy with the TNG model has a lower central density and longer cooling time compared to the one with the basic-MACER model, indicating that the AGN feedback in the TNG model is significantly stronger than in the basic-MACER model. }
\label{fig.irad}
\end{figure*}

Fig. ~\ref{fig.irad} shows the radial profiles of density, temperature, and cooling time in our two elliptical simulations. The lines coded by different colours represent the radial profiles at different times. The figure shows that the density is similar in basic-MACER and TNG outside $10\, \rm{kpc}$, but the central density  within this radius is systematically lower in TNG than in basic-MACER. Observations indicate that the central density of the hot gas in a massive elliptical galaxies is around $\sim0.1-0.4\, {\rm cm^{-3}}$ \citep{werner14}, showing that the basic-MACER model can better reproduce these density values. The temperature profiles are also similar beyond $10\, \rm{kpc}$. However, the temperature is  higher within this radius in TNG, because the cooling is stronger when the density is higher.

\citet{gaspari12} have performed idealized elliptical galaxy simulations with different AGN feedback efficiencies and obtained similar results. This finding can ultimately be understood as a consequence of the self-regulated nature of AGN feedback. Since in different AGN feedback models the AGN feedback heating should roughly balance the radiative cooling of the gaseous halo within the cooling radius -- which is comparatively large and approximately independent of the AGN feedback models -- the energy output rate in various models should be roughly the same. So we have:
\begin{equation}
\dot{E}_{\rm cool}\approx\epsilon_{1}\dot{M}_{1}c^2\approx\epsilon_{2}\dot{M}_2c^2\label{egybl},
\end{equation}
where $\dot{E}_{\rm cool}$ is the total cooling rate of the gaseous halo within the cooling radius, $\epsilon_1$ and $\epsilon_2$ are the feedback efficiencies of the two different AGN feedback models, and $\dot{M}_1$ and $\dot{M}_{2}$ are their AGN accretion rates. In both, basic-MACER and TNG, we use the Bondi-Hoyle-Lyttleton formula to estimate the accretion rate. It is proportional to the central density if we assume that the gas surrounding the black hole is regulated by the equation of state adopted for the ISM. From Equation \eqref{egybl}, we can infer that the central density must be higher for lower feedback efficiency. \citet{wellons23} reached similar insights into this phenomenon of self-regulation. Since the feedback efficiency is higher in TNG than in basic-MACER, the central density must be lower in TNG. 

This self-regulated behavior is directly illustrated in the bottom panel of Fig. \ref{fig:imdot&sfr}, which shows the AGN energy output rate as a function of time for both models. Despite the different feedback efficiencies, the AGN energy output rates in basic-MACER and TNG are comparable, consistent with Equation~\eqref{egybl}. The larger amplitude fluctuations in basic-MACER are due to its higher central gas density, which more readily gives rise to cold gas condensation near the center, episodically boosting the accretion rate.

Based on the above analysis, it is also easy to understand why no star formation appears in TNG. From Fig. \ref{fig.irad}, we can see that due to the low central density in TNG, the cooling timescale can never meet the thermal instability criterion, so that cold gas cannot form under this condition. 

The idealized elliptical galaxy simulations have helped us to understand the differences in the impact of AGN feedback in basic-MACER and TNG, especially in the low accretion rate mode. We expect that this will also help us to better understand some of the results in cosmological simulation. In general, both models can produce a quiescent elliptical galaxy, but basic-MACER will produce a galaxy with more residual activity than TNG. To more broadly understand the implications of this in other environments of galaxy formation and evolution, we need to examine the behavior of the two models in a full cosmological context, something that we turn to in the next section. 

\section{Results for cosmological simulations}\label{cosmological}

In this work, we focus on the impact of the variation of the AGN feedback model on star formation, black hole accretion, the $M_{\star}-M_{\rm{BH}}$ relation, and the gas fraction. While these are among the most important quantities of interest, it is clear that many further aspects of AGN feedback could be considered, something that would however go beyond the scope of this paper. 

\subsection{Star formation rate density}

\begin{figure}
 \includegraphics[width=\columnwidth]{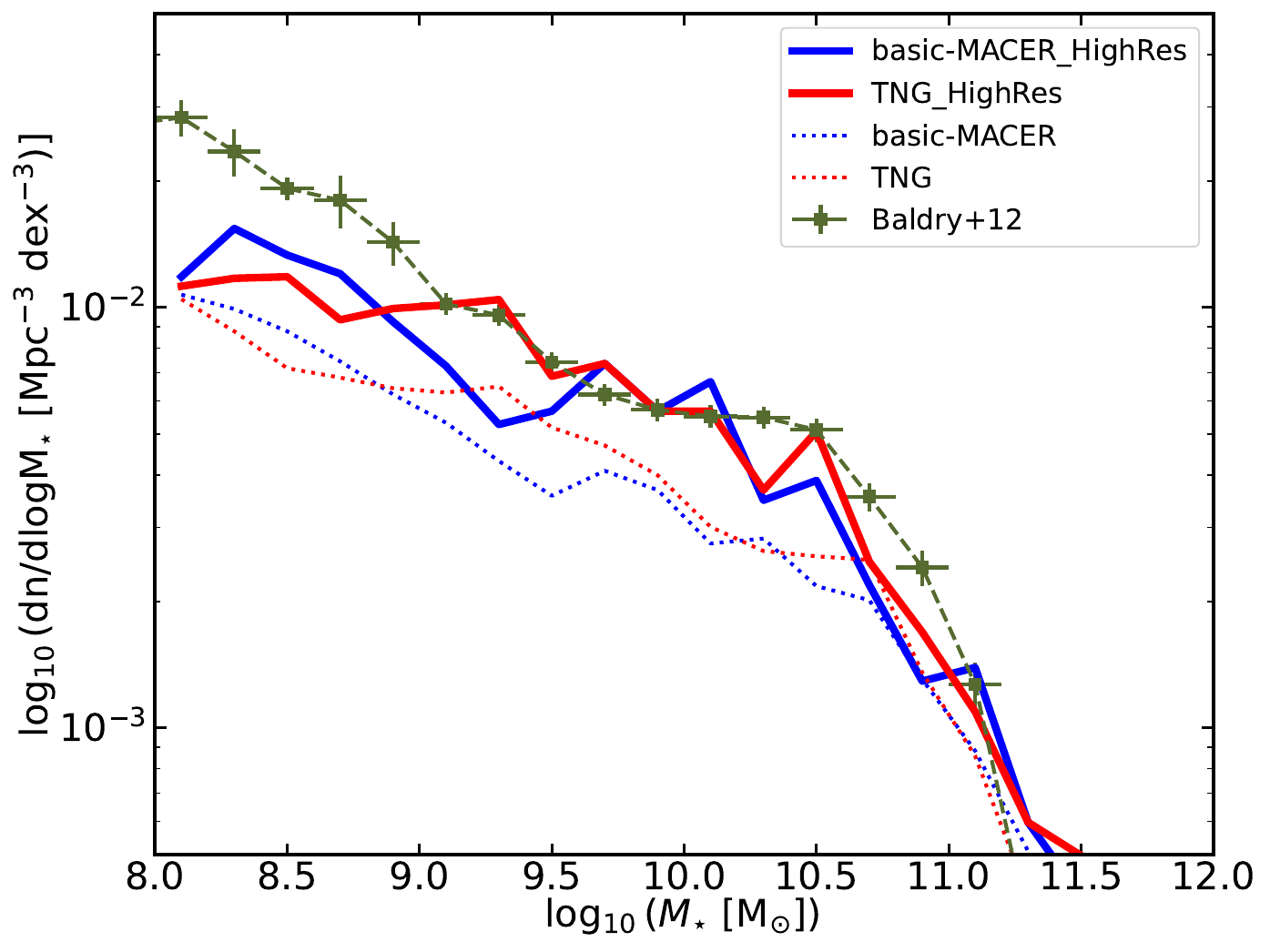}
 \caption{The galaxy stellar mass function at $z=0$ comparing the basic-MACER and TNG models across different resolutions. To illustrate the resolution dependence, higher-resolution results (basic-MACER\_HighRes and TNG\_HighRes with $2\times512^3$ particles in a $25\,h^{-1}{\rm  cMpc}$ box) are shown with thick solid lines, and standard-resolution results with dotted lines. The dark green points and dashed line represent the observational result from \citet{baldry12}, which is shown for reference.}
 \label{fig:gsmf}
\end{figure}

In Fig.~\ref{fig:gsmf}, we show the galaxy stellar mass function (GSMF) at $z=0$ for both the basic-MACER and TNG models across two different resolution levels. We find that the GSMFs predicted by the two models are very similar over the resolved stellar mass range, indicating that the modification of the AGN feedback energetics does not significantly alter the global stellar mass distribution. Observational estimates from \citet{baldry12} are shown for reference. As demonstrated by the high-resolution runs (thick solid lines), the discrepancy between the standard-resolution simulations and observations is alleviated when the resolution is increased, a trend that is fully in line with previous results for the TNG model \citep{pillepich18}. 

Since our primary focus is on evaluating the relative differences between the basic-MACER and TNG models rather than achieving absolutely converged stellar masses, and the large box size provides better statistical coverage at the massive end of the galaxy mass function, large box simulations are better suited for our purposes.

\begin{figure}
 \includegraphics[width=\columnwidth]{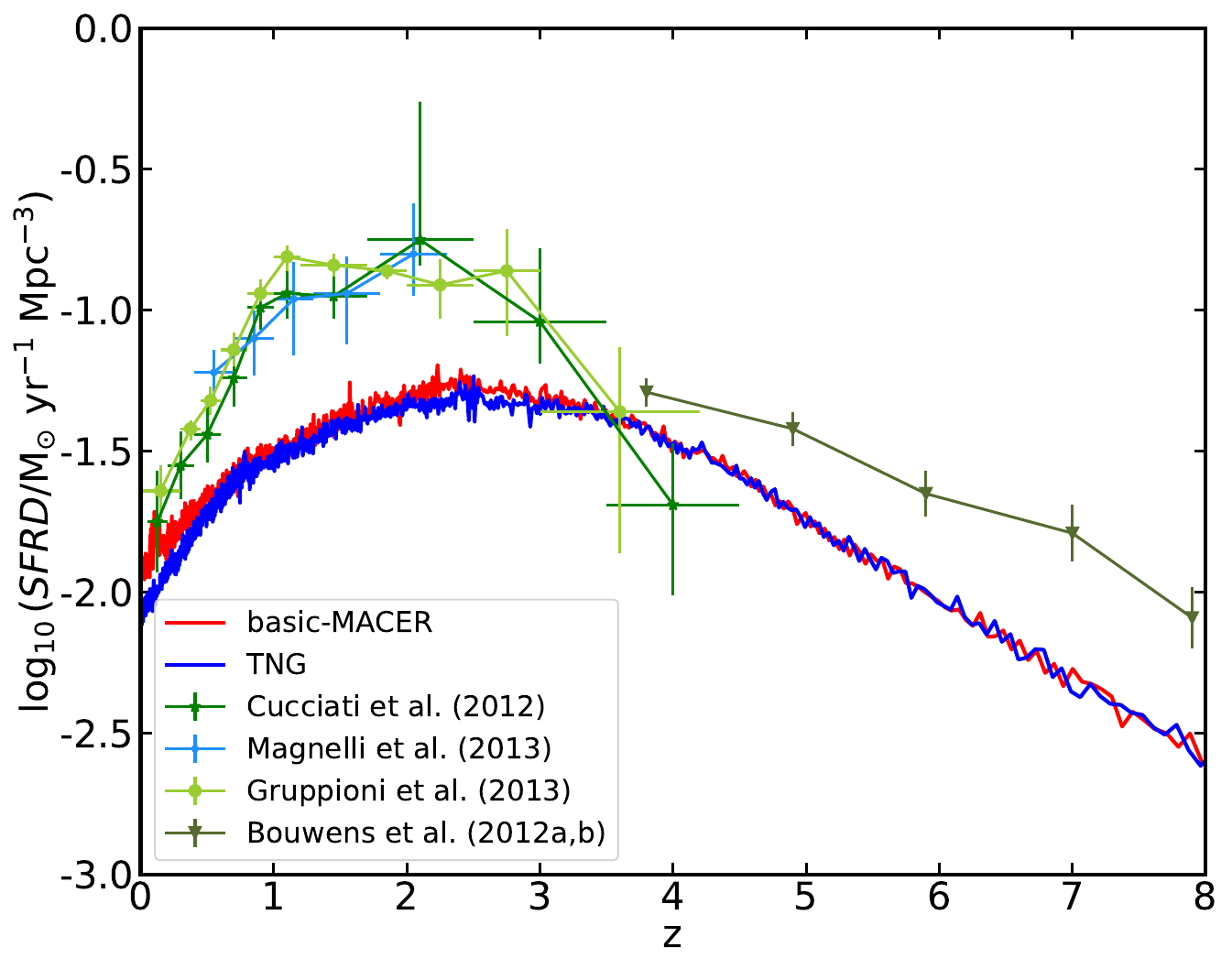}
 \caption{Star formation rate density (SFRD) as a function of redshift. The points with error bars are observational SFRD determinations from \citet{cucciati12}, \citet{magnelli13}, \citet{gruppioni13}, and \citet{bouwens14a,bouwens14b}. The SFRDs of the basic-MACER and TNG simulations agree quite well with each other and are broadly consistent with the data. There is a more pronounced late-time reduction of the SFRD in the TNG model, reflecting the overall stronger impact of its AGN feedback model.}
 \label{fig:cssfrd}
\end{figure}

In Fig.~\ref{fig:cssfrd}, we show the cosmic star formation rate density (SFRD) as a function of redshift $z$ for the two AGN feedback models. We can see that the overall result from the basic-MACER model is similar to that from the TNG model except for redshifts $z \le 1$, where there is clearly a slight deviation between the two simulations. 

\begin{figure}
 \includegraphics[width=\columnwidth]{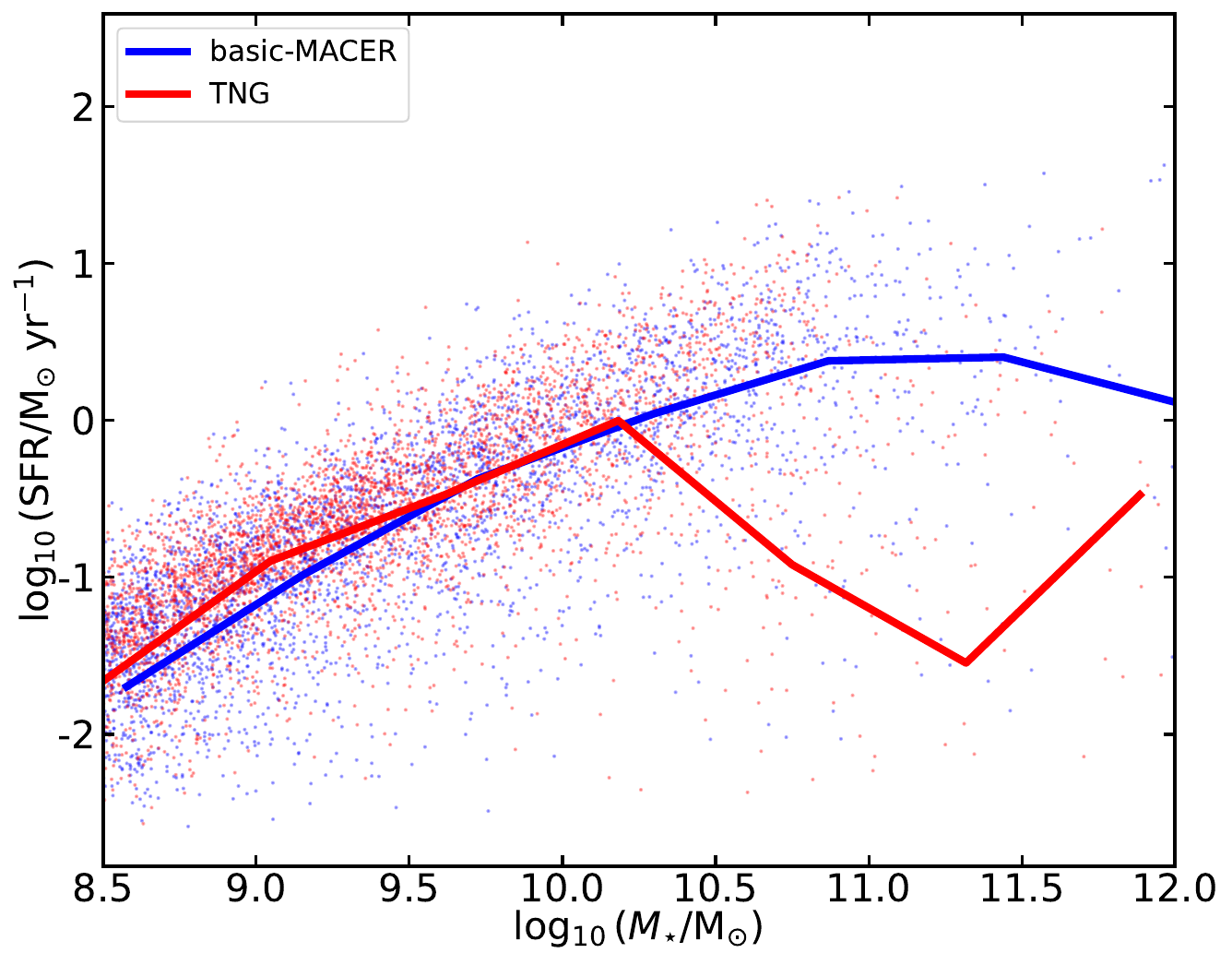}
 \caption{Relation between SFR and stellar mass  at $z=0$ in the cosmological simulations with the basic-MACER and TNG models. Each point represents a single galaxy, and the lines denote a running median SFR as a function of stellar mass. The figure demonstrates that while the median SFR of the galaxies for $M_{\star}<10\,{\rm M_{\odot}}$ is similar, there is a relatively sharply defined transition mass scale in TNG where strongly quenched galaxies appear, whereas the quenching of massive galaxies sets in more gradually in the basic-MACER model.}
 \label{fig:sfrvstar}
\end{figure}

We further investigate the SFR for different stellar masses. Fig.~\ref{fig:sfrvstar} shows the SFR as a function of stellar mass at $z=0$. We can see that both the TNG and basic-MACER models can reproduce a reasonable star formation main sequence. However, for stellar masses larger than $10^{10}\,{\rm M_{\odot}}$, there are fewer quiescent galaxies in basic-MACER. This result implies that the AGN feedback in basic-MACER is less efficient in quenching the massive galaxy. This finding can be explained by the idealized elliptical galaxy simulations. From the results of idealized simulations in Fig.~\ref{fig.irad}, we can see that the central density is much higher in basic-MACER than in TNG because of lower feedback efficiency. 

\begin{figure*}
 \includegraphics[width=\linewidth]{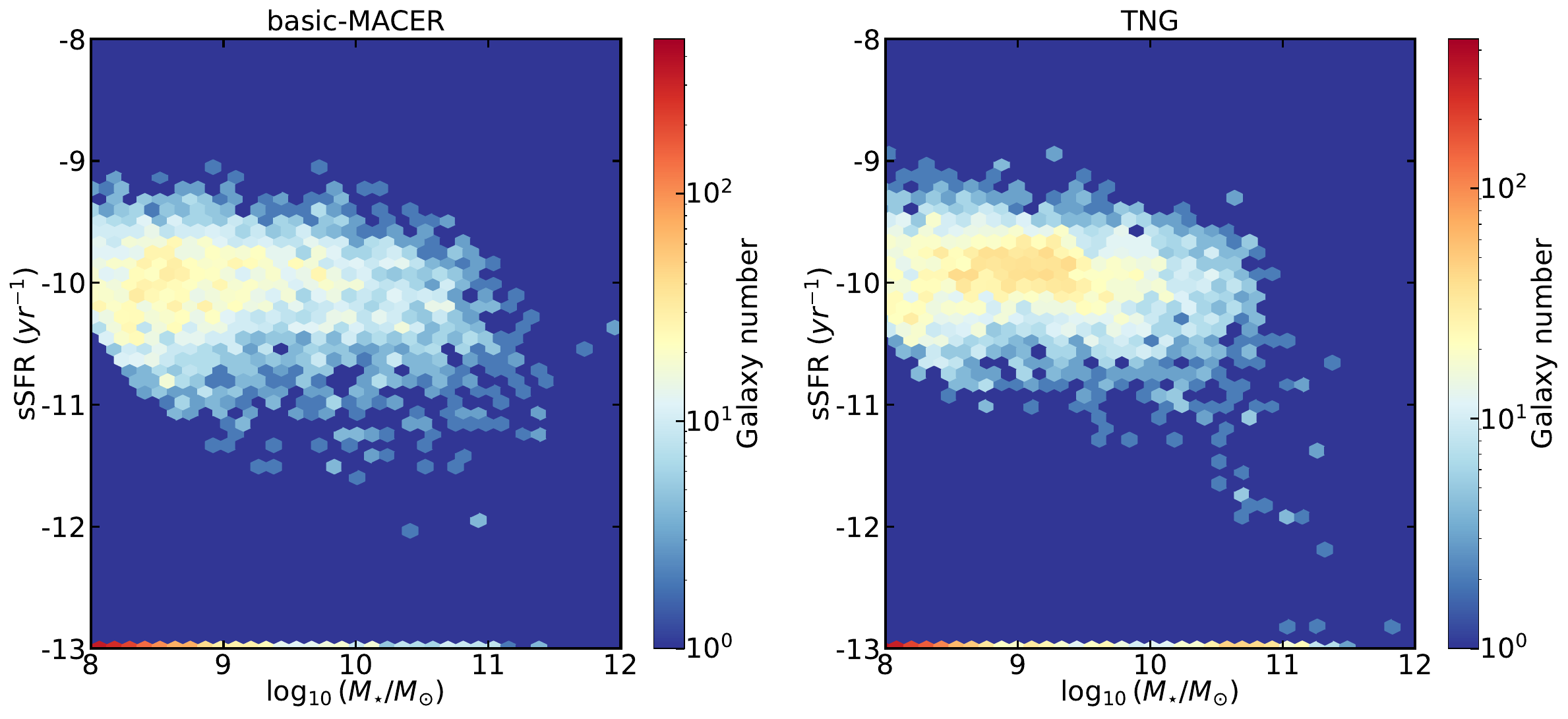}
 \caption{Two-dimensional histogram of simulated galaxies in the basic-MACER and TNG models at $z=0$ on the specific SFR vs stellar mass plane. The figure shows that there are more quenched massive galaxies in TNG than in basic-MACER. 
 }
 \label{fig:sSFRdis}
\end{figure*}

This trend can be seen more clearly in the specific star formation rate (sSFR) distribution. Fig.~\ref{fig:sSFRdis} shows the sSFR distribution at $z=0$ in the sSFR-stellar mass plane. TNG can reproduce a good bimodality in the sSFR-$M_{\star}$ diagram \citep{nelson18}. However, it produces too many over-quenched massive galaxies without any star formation. Compared with TNG, galaxies more massive than $10^{10}\,{\rm M_{\odot}}$ are more star-forming in basic-MACER. 

\begin{figure*}
 \includegraphics[width=\linewidth]{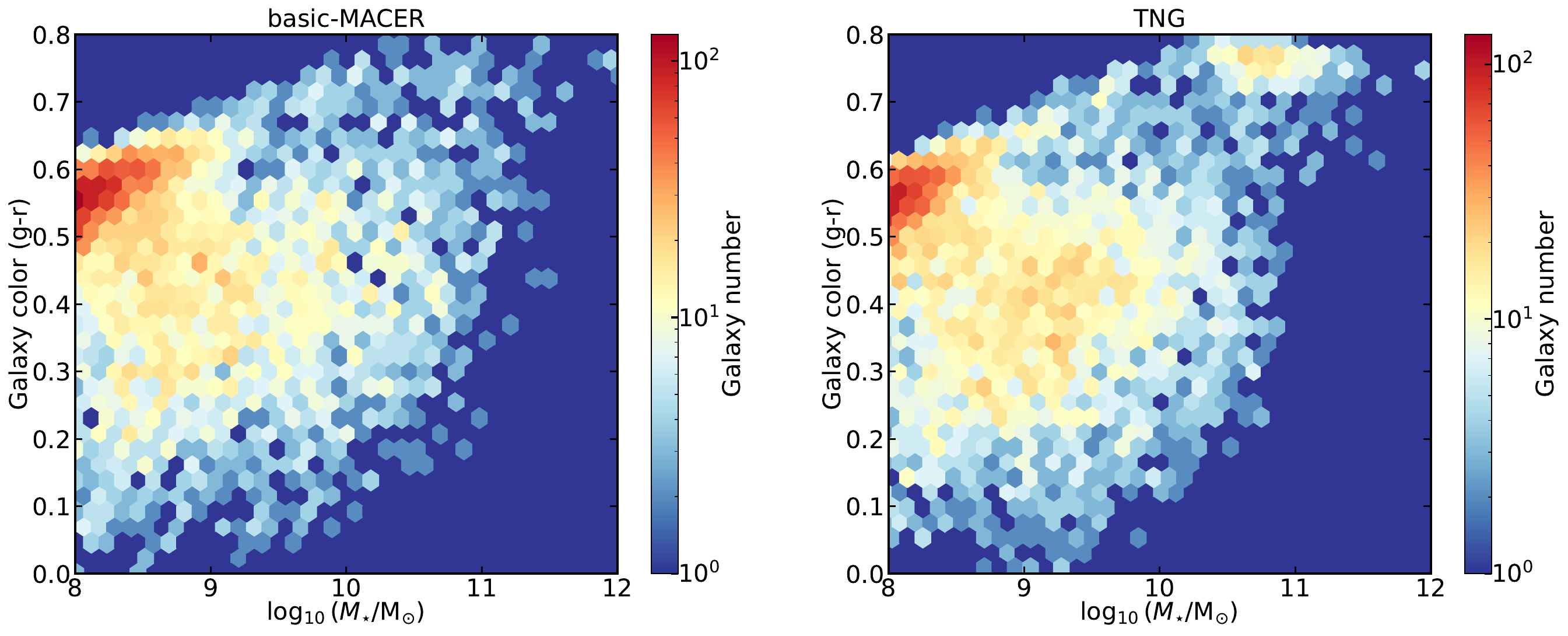}
 \caption{Two-dimensional histogram of simulated galaxis in the basic-MACER and TNG models at $z=0$ on the g-r colour vs stellar mass plane. The simulation with the TNG model (right) shows a pronounced colour bimodality that is largely absent in basic-MACER (left).}
 \label{fig:colorbi}
\end{figure*}

Fig.~\ref{fig:colorbi} shows the colour bimodality at $z=0$ for the two models. As found in \citet{nelson18}, TNG produces a reasonably good colour bimodality consistent with observations. The galaxy colour can be divided into two main subclasses. Massive galaxies with $M_{\star}\geq 10^{11}\,{\rm M_{\odot}}$ reside in the red region with $g-r\sim0.8$, while less massive galaxies mainly reside in the blue region with $g-r\sim 0.4$. However, the bimodality is weaker in basic-MACER than in TNG. Since we already found that in basic-MACER it is harder to fully quench massive galaxies than TNG, it is obvious that young stars will at a higher level contribute to the light in basic-MACER massive galaxies. This is why the colour is typically bluer in basic-MACER than in TNG.

\subsection{Black holes}

Correlations between the  black hole and stellar masses of the  host galaxies are believed to provide strong evidence for a co-evolution between black holes and their host galaxies. The key physical mechanism for establishing this co-evolution is thought to be AGN feedback. In this subsection, we thus focus on analyzing differences in the $M_{\star}-M_{\rm{BH}}$ relation in the two different AGN feedback models. 

\begin{figure*}
 \includegraphics[width=\linewidth]{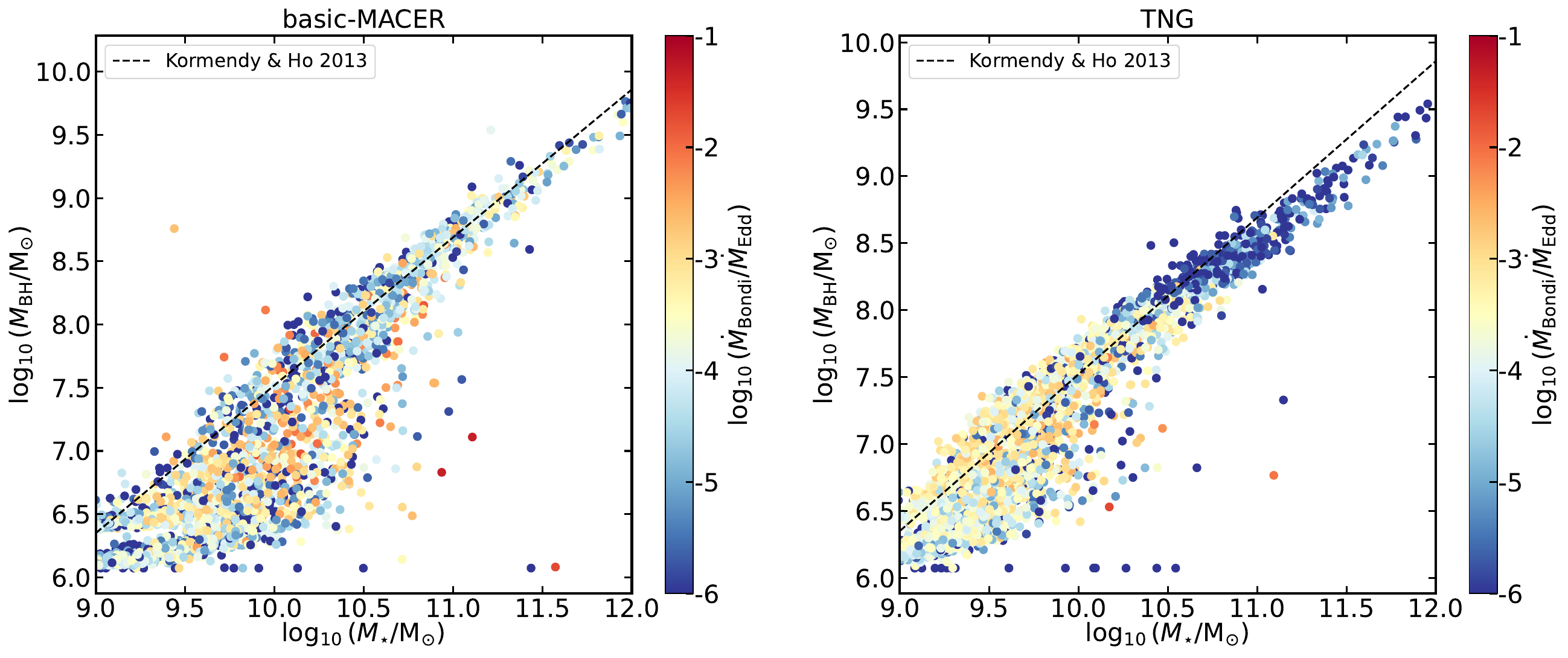}
 \caption{The relation between stellar mass and black hole mass predicted at $z=0$ by the basic-MACER and TNG models. The colour of the points encodes the value of the Bondi accretion rate. The simulation with the TNG model shows a relatively sharply defined transition of the Bondi accretion rates towards low values at $M_{\rm BH}\sim10^{8}\,{\rm M_{\odot}}$, while the simulation with the basic-MACER model does not show such a feature.}
 \label{fig:mbhms}
\end{figure*}

\begin{figure}
 \includegraphics[width=\linewidth]{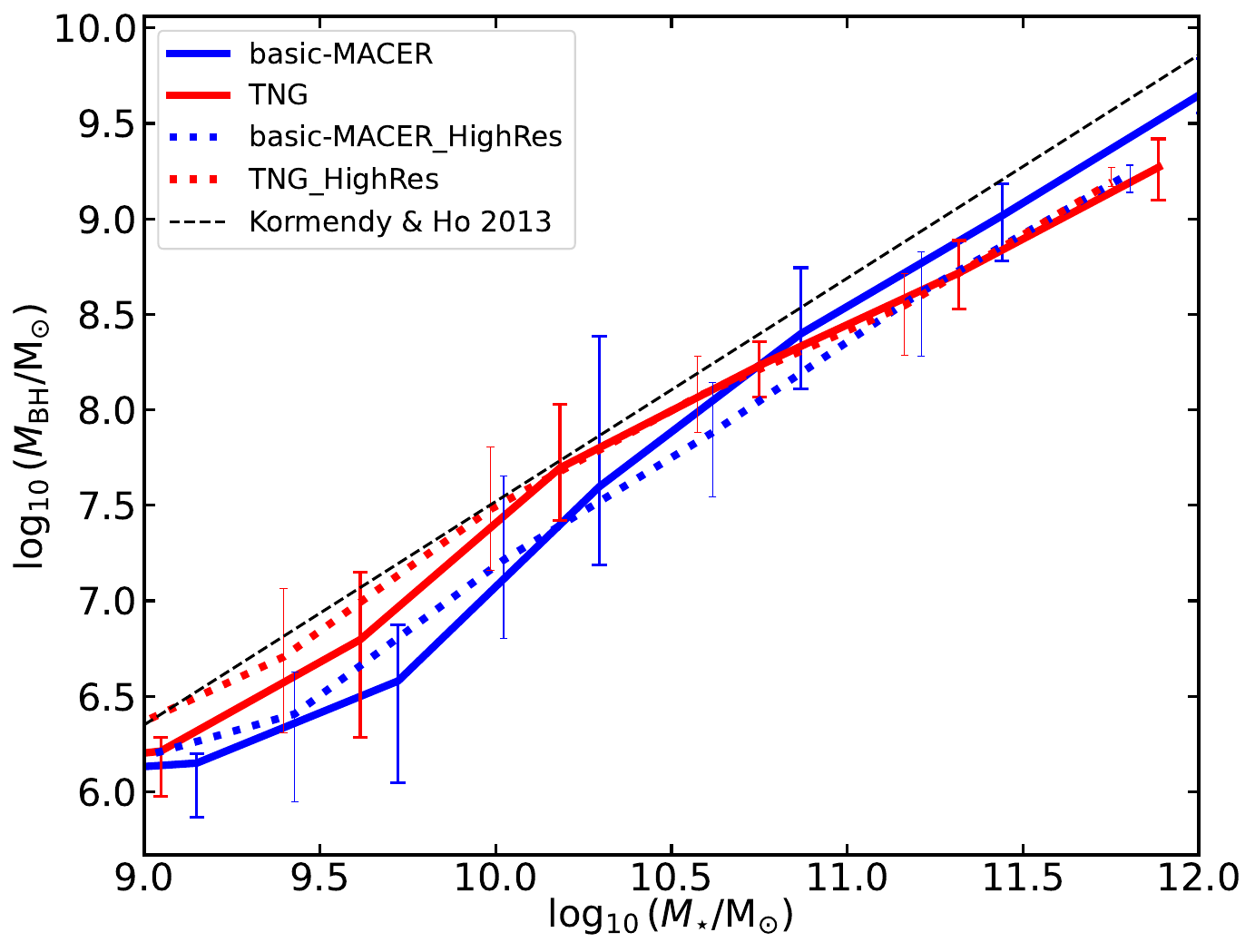}
\caption{The $M_\star$-$M_{\rm BH}$ relation at $z=0$ in the basic-MACER and TNG models at two different resolutions. Solid lines show the fiducial runs (basic-MACER and TNG); dotted lines show the higher-resolution runs (basic-MACER\_HighRes and TNG\_HighRes). Error bars indicate the 16th--84th percentiles. The black dashed line is the best-fitting $M_{\rm Bulge}$--$M_{\rm BH}$ relation from \citet{kormendy13}, shown for reference. The global trend of the relation is roughly converged, while both models show a modest upward shift at lower stellar masses when the resolution is increased.}
\label{fig:mstar_mbh_res}
\end{figure}

Fig.~\ref{fig:mbhms} shows the predicted $M_{\star}-M_{\rm{BH}}$ relation for the different models. The colour encodes the Eddington ratio attained for the Bondi accretion rate. The dotted line represents the best-fitting $M_{\rm{Bulge}}-M_{\rm{BH}}$ relation from \citet{kormendy13}. Since the relation from \citet{kormendy13} is the correlation between black hole mass and bulge mass instead of stellar mass, and the bulge does not dominate the stellar mass in low-mass galaxies, it just provides a reference for low-mass systems. From the figure, we can see that both TNG and basic-MACER can reproduce the scaling relation for massive galaxies reasonably well, which arises here because AGN feedback can suppress the further growth of the central black holes. 

We have verified that these results are not sensitive to numerical resolution. Fig.~\ref{fig:mstar_mbh_res} compares the $M_\star$-$M_{\rm BH}$ relation in our fiducial runs with the higher-resolution runs (basic-MACER\_HighRes and TNG\_HighRes). The relation is roughly converged. At lower stellar masses, both models show a slight upward shift at higher resolution, but the global trend is unaffected.

\begin{figure}
 \includegraphics[width=\linewidth]{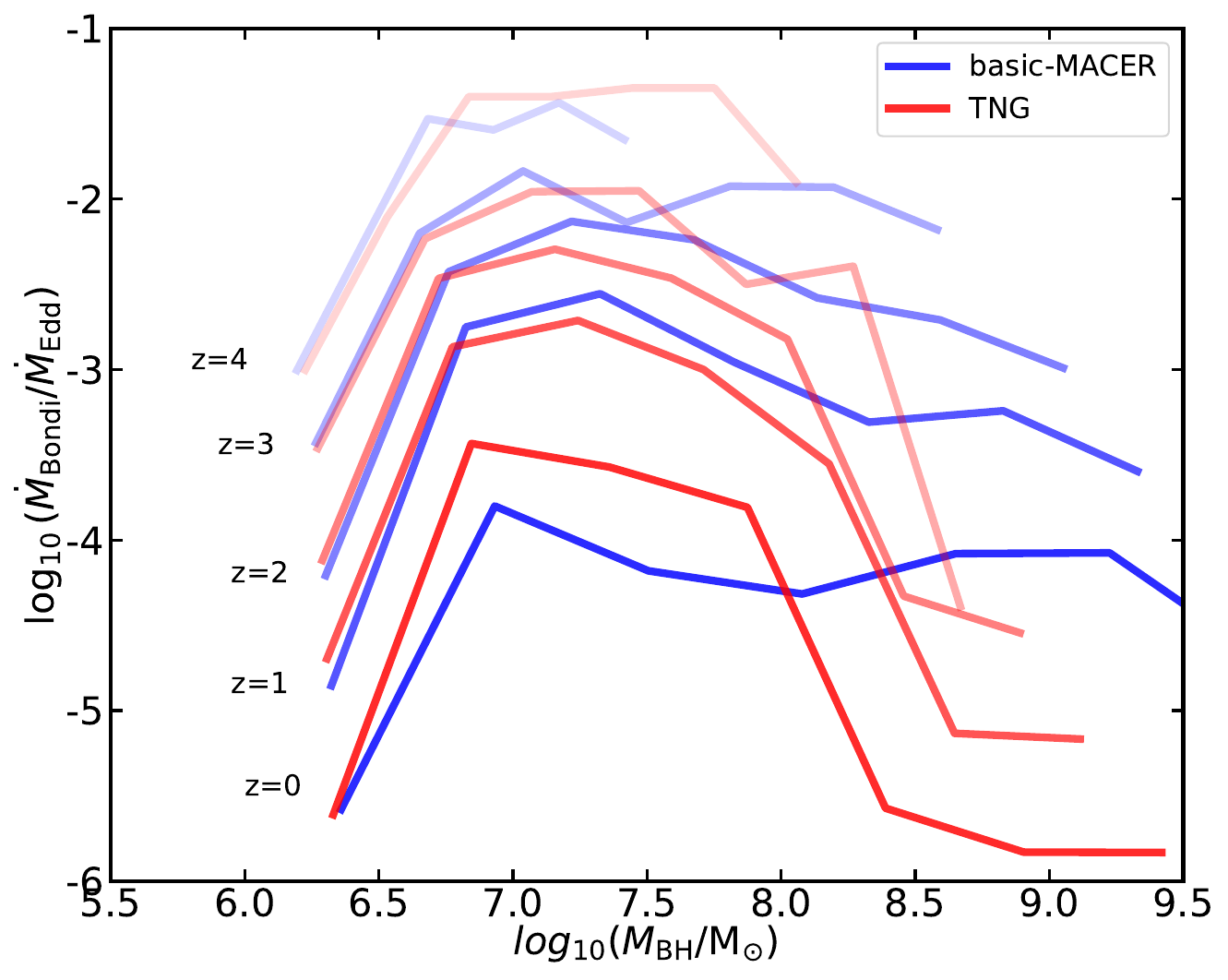}
 \caption{The median Bondi accretion rate as a function of black hole mass at redshifts $z=4$, $3$, $2$, $1$, and $0$, in the basic-MACER and TNG models. The Bondi accretion rate in the simulation with the TNG model shows a significant drop at $M_{\rm BH}\sim10^{8}\,{\rm M_{\odot}}$, while the basic-MACER model does not have a similar trend.}
 \label{fig:mdotb}
\end{figure}

\begin{figure}
 \includegraphics[width=\linewidth]{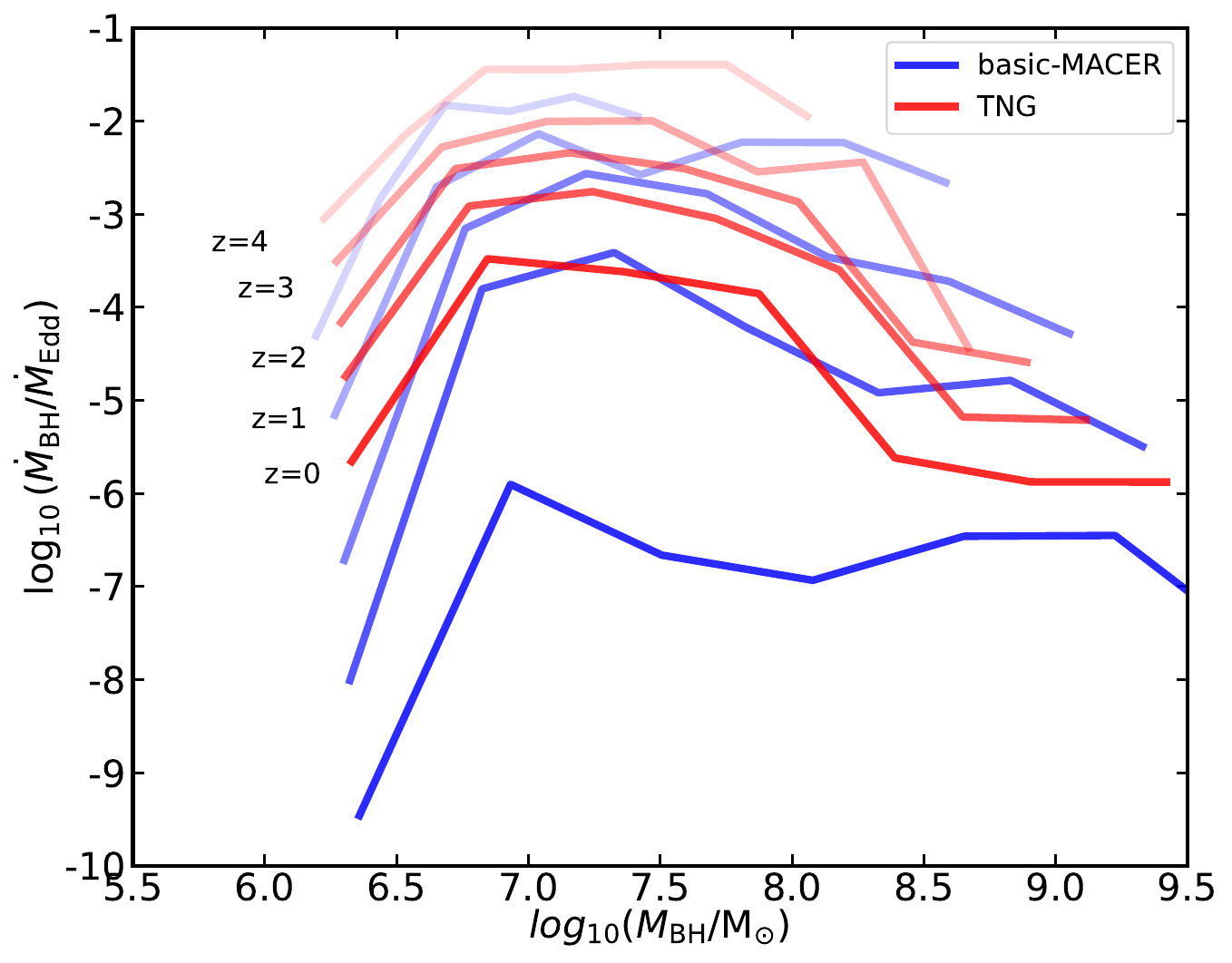}
 \caption{The median BHAR as a function of black hole mass at redshifts $z=4$, $3$, $2$, $1$, and $0$, in basic-MACER and TNG. In contrast to Fig.~\ref{fig:mdotb}, we here show the ``real'' BH accretion rate, which ends up being  often lower in basic-MACER than in TNG due to its different assumptions about the accretion fraction.}
 \label{fig:mdotbh}
\end{figure}

However, TNG produces slightly over-massive black holes for low-mass galaxies compared to basic-MACER. This difference is due to the lower accretion fraction in the latter. Fig.~\ref{fig:mdotb} and Fig.~\ref{fig:mdotbh} show the median Bondi accretion rate and the median BHAR as a function of black hole mass at different redshift values. Comparing the two figures, we can see that the Bondi accretion rates in both models are similar for low-mass black holes. However, the BHAR is significantly lower in basic-MACER for these black hole masses nevertheless. For black holes with similar Bondi accretion rates, the actual growth can still be different due to different accretion fractions, explaining why black holes grow more slowly in the basic-MACER model. And this discrepancy in the black hole growth between the two models tends to become larger when the black holes grow. Note that the mass of the black hole seeds in the two simulations is equal and around $\sim10^6\,{\rm M_{\odot}}$. If they grow with the Eddington accretion rate, but have different accretion fractions of $0.5$ and $0.9$, their mass ratio can become larger than a factor of 10 after a time of $3\times10^8 \,\rm{yr}$. This thus explains why the black hole masses in basic-MACER are smaller than in TNG, especially at the lower mass end. 

Focusing on the BHAR shown in Fig.~\ref{fig:mdotbh}, we find that there is a steep decline with increasing black hole mass around $10^{8}\,{\rm M_{\odot}}$, which corresponds to a stellar mass around $10^{10}\,{\rm M_{\odot}}$ in the TNG model. Previous works have found that this occcurs due  to the transition to the much more efficient mode of feedback at low accretion rates  \citep{weinberger17}. The corresponding transition accretion rate is scaled  with black hole mass in the TNG model, making it easier for large black holes to reach it. As black holes grow, they will therefore  gradually become ever more dominated by the radio mode kinetic feedback. Because kinetic feedback is stronger than thermal feedback at a high accretion rate, the black hole accretion then becomes strongly suppressed. 

In contrast, the Bondi accretion rate does not have a strong, built-in decline in basic-MACER, but rather changes smoothly with black hole mass. This result is due to a relatively smooth transition of the accretion rate in basic-MACER. The energy output in both cold and hot feedback modes in basic-MACER is dominated by the kinetic energy of the wind/jet, and the transition accretion rate is independent of black hole mass.

\subsection{Gas fractions and gaseous profiles}

The  AGN feedback will also affect the gas properties in the halo. In this subsection, we therefore investigate the gas fractions, the  radial gas profiles, and the X-ray luminosities from the cosmological simulations with the two different feedback models.

\begin{figure*}
 \includegraphics[width=\linewidth]{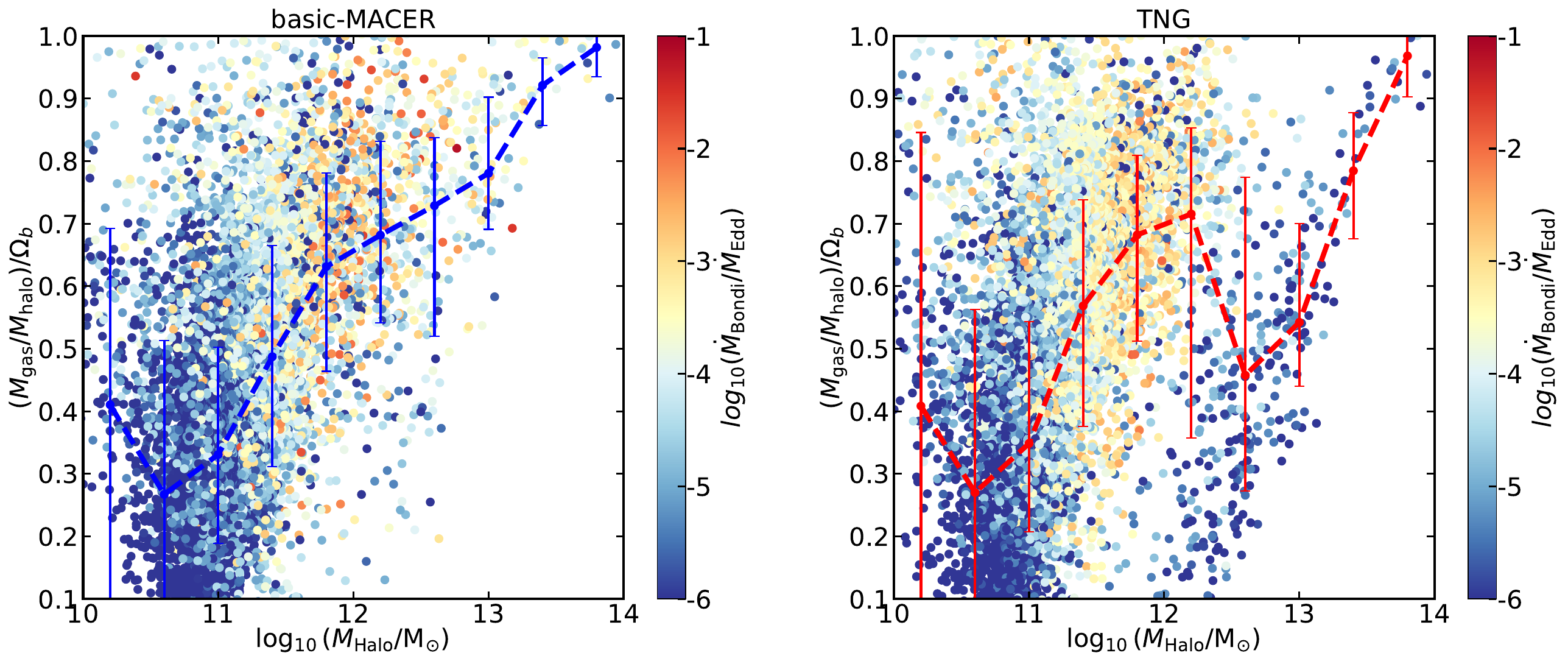}
 \caption{Scatter plot of gas mass fraction versus halo mass of the simulated galaxies in the basic-MACER and TNG models at $z=0$. The colour of the points encodes the Bondi accretion rate. The solid lines show the median gas fraction in bins of halo mass, and the error bars indicate the 16th–84th percentiles. The simulation with the basic-MACER model shows a continuous distribution of the baryonic gas fraction, while the simulation with the TNG model exhibits a significant drop at $M_{\rm halo}\sim10^{12}\, {\rm M_{\odot}}$ when the kinetic AGN feedback sets in.}
 \label{fig:gas}
\end{figure*}

Fig.~\ref{fig:gas} shows the gas mass fraction as a function of halo mass for the two AGN feedback models. The circles are colour-coded by the Eddington ratio of the Bondi accretion rate. The lines represent the median gas fraction in bins of halo mass, and the error bars indicate the 16th–84th percentiles. The figure shows that the gas fraction is similar for the two models when the halo mass is smaller than $10^{12}\,{\rm M_{\odot}}$, but for halo mass above $10^{12}\,{\rm M_{\odot}}$,  the gas fraction is significantly lower in TNG than in basic-MACER. Referring to the corresponding Bondi accretion rate, we can see that the Bondi accretion rate in TNG exhibits a sudden transition at $M_{\rm halo}\approx 10^{12}\,{\rm M_{\odot}}$. For larger halo masses, the Bondi accretion rate quickly drops below $10^{-4}\dot{M}_{\rm{Edd}}$, and the black holes are dominated by kinetic feedback. This result indicates that kinetic feedback can efficiently reduce the total amount of gas in the halo and produces a low gas fraction in massive halos in TNG. The gas fraction and the Bondi accretion rate are higher in basic-MACER than in TNG. This result underlines that AGN feedback in massive halos is  weaker in basic-MACER than in TNG. The feedback energy in basic-MACER is not reaching scales equally large, and consequently the whole gaseous halos of massive halos are affected less.

\section{Parameter study}\label{sec:param}

In order to better understand the universality of the differences between the basic-MACER and TNG models, we have run additional smaller box simulations that varied some of the parameters of basic-MACER. The numerical resolution of these simulations is the same as for the cosmological simulations described above, but with a $25~{\rm Mpc}/h$ box size and $2\times256^3$ particles, their volume is 8 times smaller (see Table~\ref{table:des} for a summary of all model variants and Table~\ref{table:run} for the corresponding simulation setup).

\subsection{Energy output form for high accretion rate}

One of the most significant differences between basic-MACER and TNG is the energy released by the high accretion rate. In TNG, the energy output is injected as thermal energy based on the notion that a good fraction of the feedback energy released by the BH will thermalize in any case, and in this way will couple to the surrounding ISM. In contrast, the energy output in the basic-MACER model is assumed to be in kinetic form  and to mimic observed AGN outflows like in UFOs or BAL wind systems. To study the effects of variations in the energy from the high accretion rate state, we have run a test simulation with thermal energy injection but with the same feedback efficiency as in basic-MACER (ThermalHigh;see Table~\ref{table:des}).

\begin{figure}
 \includegraphics[width=\linewidth]{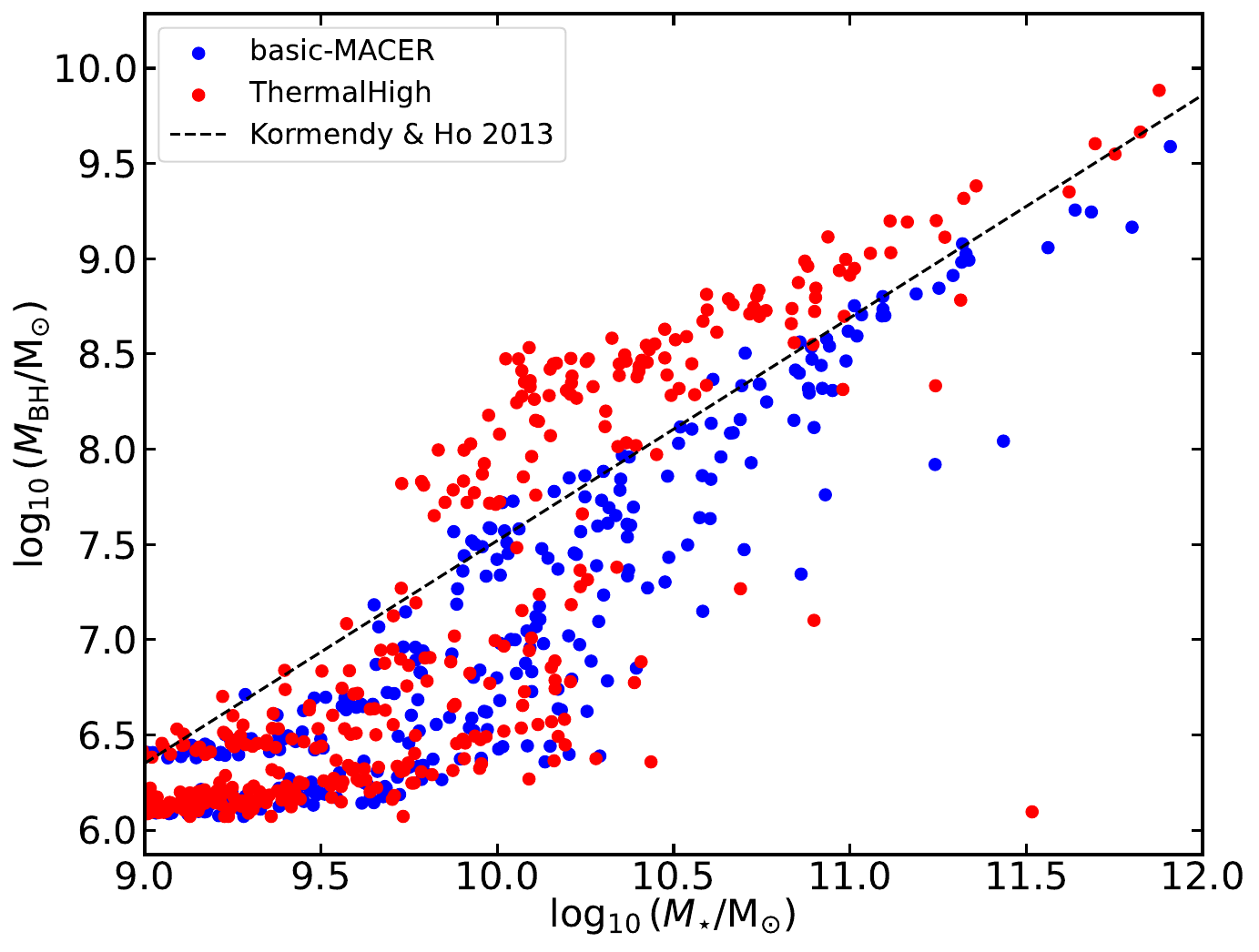}
 \caption{The black hole mass versus stellar mass relation, comparing the   fiducial basic-MACER model and the ThermalHigh variant at $z=0$. The ThermalHigh model can be viewed as a hydrid between the TNG quasar mode, a basic-MACER radio mode, and a transition accretion rate as in the basic-MACER model. The figure shows that with a thermal injection of the AGN feedback in the quasar mode, low-mass BHs can grow faster than in a simulation with the fiducial basic-MACER model.}
 \label{fig:ps1mbhms}
\end{figure}

Fig.~\ref{fig:ps1mbhms} shows the $M_{\star}-M_{\rm{BH}}$ relation at $z=0$ for the basic-MACER and the ThermalHigh models. Surprisingly, compared with basic-MACER, ThermalHigh produces a $M_{\star}-M_{\rm{BH}}$ relation with two different distinct regions. For galaxies with stellar mass less than $10^{10}\,{\rm M_{\odot}}$, almost all of the black hole mass resides around $10^{6}M_{\odot}$. Since the black hole seed is set to $\sim10^6\,{\rm M_{\odot}}$, it is clear that black holes nearly do not grow in this regime. While for galaxies with a stellar mass larger than $10^{10}\,{\rm M_{\odot}}$, most of the black holes reside systematically above the observed $M_{\rm{BH}}-M_{\rm{Bulge}}$ relation,  but with a similar slope. This result implies that black holes must have entered a fast-growth period in ThermalHigh at stellar masses larger than $10^{10}\,{\rm M_{\odot}}$, while such a fast-growth period does not exist in the basic-MACER model in this mass range.

\begin{figure*}
 \includegraphics[width=\linewidth]{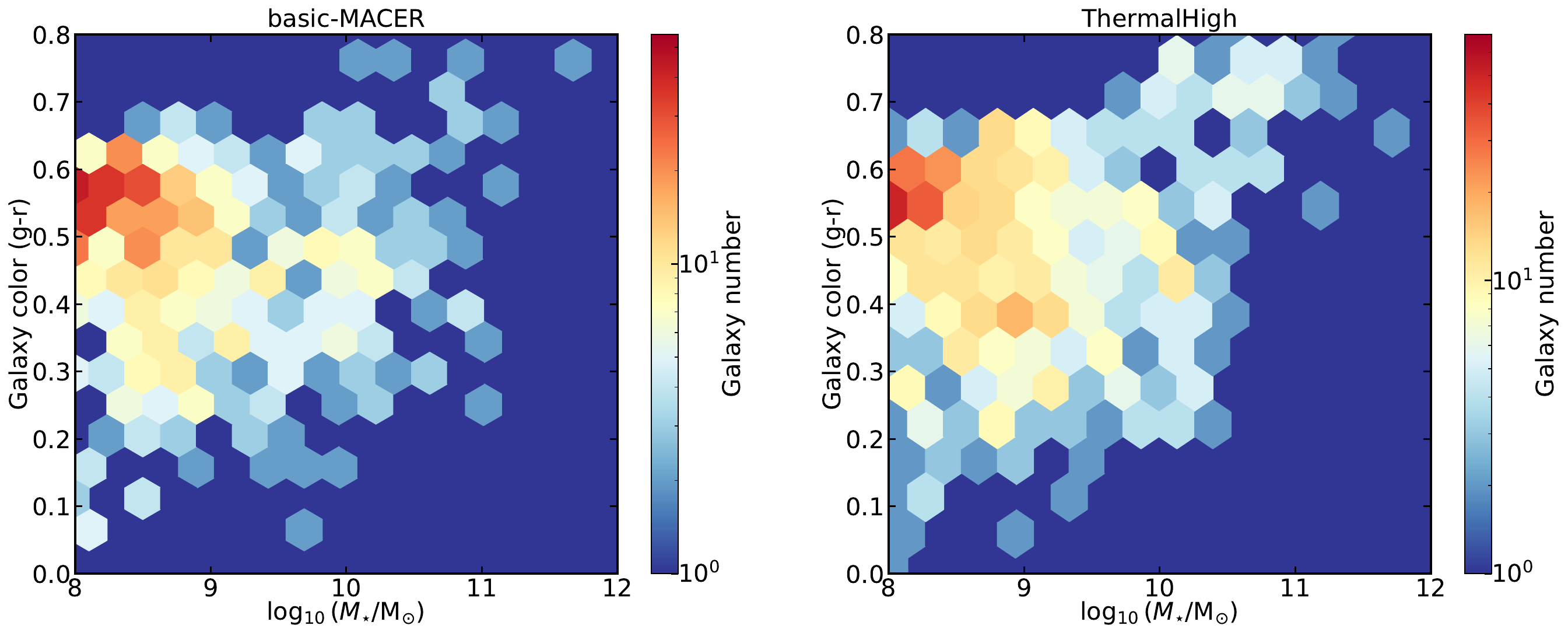}
 \caption{The colour bimodality at $z=0$ in basic-MACER and ThermalHigh. With a thermal injection for the quasar mode of the basic-MACER model. Incorporating thermal injection into the quasar mode of the basic-MACER model appears to enhance this bimodality compared to the fiducial simulation, albeit marginally.
 }
 \label{fig:ps1color}
\end{figure*}

Fig.~\ref{fig:ps1color} shows the colour bimodality at $z=0$ for basic-MACER and ThermalHigh. The bimodality is more pronounced in ThermalHigh, albeit marginally. The larger black hole mass can explain this result at $M_{\star}\sim10^{10}\,{\rm M_{\odot}}$. With a larger black hole mass, the released feedback energy in ThermalHigh is higher than in basic-MACER, and can more  efficiently prevent the cooling of the gaseous halo. This result also reminds us that AGN feedback at galaxy stellar masses around $10^{10}\,{\rm M_{\odot}}$ is particularly important for producing a correct $M_{\star}-M_{\rm{BH}}$ relation and colour bimodality. We will further investigate this issue in forthcoming work.

\subsection{Variations of the feedback efficiency at high accretion rate}

\begin{figure}
 \includegraphics[width=\linewidth]{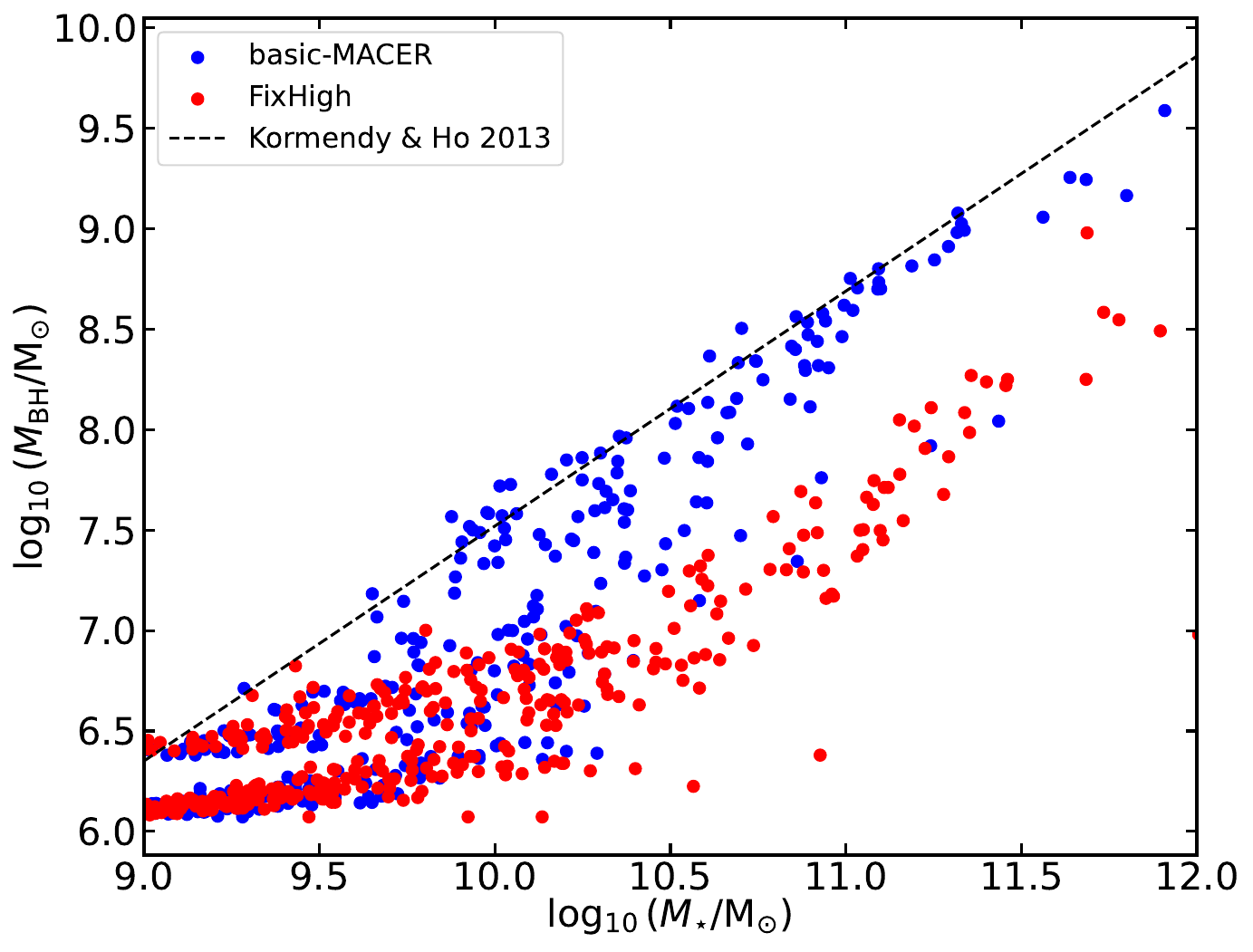}
 \caption{The black hole mass and stellar mass relation produced from  basic-MACER and the FixHigh model variant at $z=0$, where the FixHigh run is set to a constant feedback efficiency for different BH masses in the MACER model. The figure shows that BHs do  not grow their mass efficiently in the FixHigh run, indicating that low-mass BHs should have a weaker quasar wind feedback than massive BHs, unlike what FixHigh implements.}
 \label{fig:ps2mbhms}
\end{figure}

In addition to the different energy forms at high accretion rates, basic-MACER and TNG also have different feedback efficiency. The feedback efficiency at high accretion rate in basic-MACER changes with black holes mass. To investigate the impacts of this feature, we run a test simulation (FixHigh) with the feedback efficiency fixed to the value in basic-MACER at a black hole mass of $10^{8}\,{\rm M}_{\odot}$. This selection makes the feedback efficiency higher than in the basic-MACER model for smaller black holes masses, and lower for more massive black holes. 

Fig.~\ref{fig:ps2mbhms} shows the $M_{\star}-M_{\rm{BH}}$ relation produced from FixHigh simulation, which we find to systematically deviate from observations. Further investigating the $M_{\rm{BH}}-M_{\rm{halo}}$ and $M_{\star}-M_{\rm{halo}}$ relations, we find that the stellar masses in the simulations do not exhibit significant differences. The deviation in the $M_{\star}-M_{\rm{BH}}$ relation is instead mainly caused by black hole mass differences resulting from an inefficient black hole growth. This result implies that the kinetic feedback efficiency at high accretion rates has an important influence on the growth of black holes. A high feedback efficiency for smaller black holes leads to inefficient black hole growth and produces the discrepancy.

Observations have found that the metallicities of AGN outflows increase with increasing black hole mass \citep{wang12, wang22}. The dependence of AGN outflow strength on black hole mass is constrained by observations \citep[e.g., ][]{perna17}. In particular, it is widely accepted that radiation pressure from spectral lines (hereafter line forces) is one of the main launching mechanism for  AGN outflows \citep[e.g.,][]{proga2000}. Since the line force strength depends on metallicity, the strength of AGN outflows is naturally correlated with black hole mass. 

The variation of feedback efficiency with black hole mass ensures that in the basic-MACER model the transition accretion rate does not need to be changed explicitly  with black hole mass. One of the motivations for introducing this in TNG was to prevent black holes from being dominated by kinetic feedback when they are still small, which would  greatly suppress the growth of the black hole. In basic-MACER, when the black hole mass is not yet  large, the AGN feedback is also weak in the quasar mode. The black hole can then grow successfully, allowing the model to match the observed $M_{\star}-M_{\rm{BH}}$ relation.

\subsection{The efficiency at low accretion rates}

One of the differences between basic-MACER and TNG is the feedback efficiency at low accretion rates. From the upper right panel of Fig.~\ref{fig:mmodel}, we can see that at low accretion rates, the feedback efficiency in basic-MACER is lower than in TNG, even including the jet energy output in the former. From the elliptical galaxy simulations, we already got some insight into the importance of the radio mode in keeping elliptical galaxies quiescent, as black holes spend most of their time in the low accretion rate regime.  In this section, we further investigate this effect by considering variations of the corresponding feedback efficiency.

\begin{figure*}
 \includegraphics[width=\linewidth]{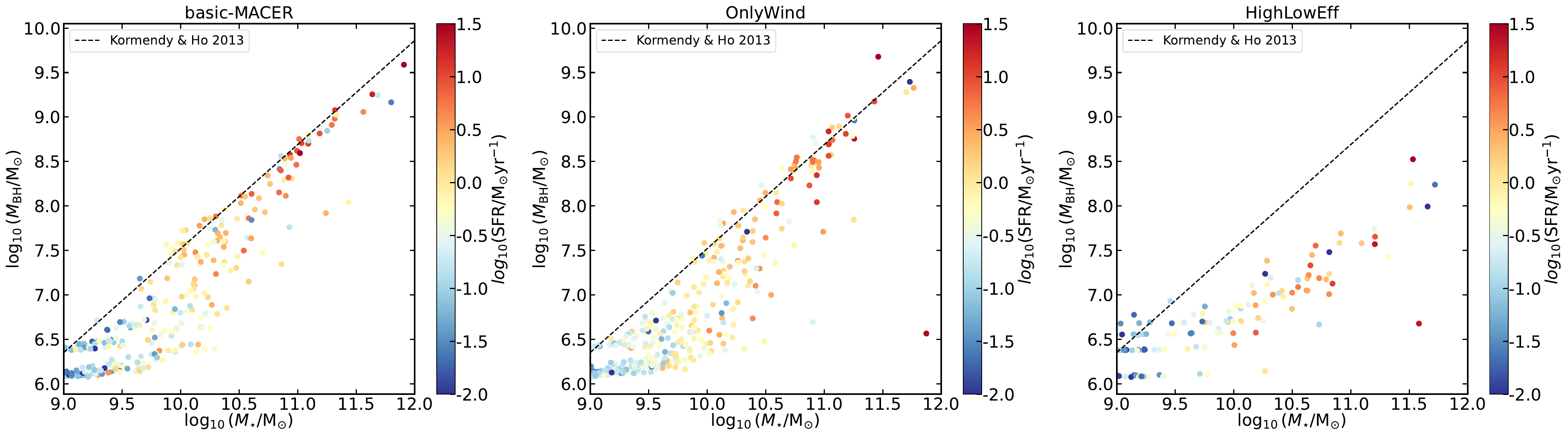}
 \caption{The black hole mass versus stellar mass relations produced by the  basic-MACER, OnlyWind and HighLowEff simulations at $z=0$. The colour of the points encodes the SFR. Here the OnlyWind model variant ignores the jet energy that is included in  the radio mode of the basic-MACER model, whereas the HighLowEff model corresponds to basic-MACER but replaces the radio mode with that of the TNG model. The figure shows that the jet energy has minor effects on the BH growth, but the radio mode of the TNG model very strongly suppresses the BH growth.}
 \label{fig:ps3mbhms}
\end{figure*}

We use three cosmological simulation tests to study the corresponding effects. The first test calculation only considers the feedback energy provided by the kinetic wind but neglects the jet (OnlyWind), the second is the  basic-MACER model, while in the third the feedback efficiency is set to 0.2, the same value as in TNG (HighLowEff).  Fig.~\ref{fig:ps3mbhms} shows the $M_{\star}-M_{\rm{BH}}$ relation produced by the three different models. The points are colour-coded by the SFR. For the models with and without a jet, the results do not show significant differences except that the SFR in massive galaxies is slightly stronger in the simulation without a jet. Since the first test simulation does not include the jet energy, the kinetic energy output for massive galaxies is weaker, and the massive galaxies in this simulation are bound to become more active, and naturally, their SFR will be higher. 

However, when the feedback efficiency becomes larger and is set to 0.2, the $M_{\star}-M_{\rm{BH}}$ relation becomes significantly different from that of the other two simulations. From Fig.~\ref{fig:ps3mbhms}, we can see that the black holes cannot reach $10^{8}\,{\rm M_{\odot}}$ in this case, and the black hole growth is strongly suppressed. The resulting $M_{\star}-M_{\rm{BH}}$ relation deviates from observations significantly.

The results can also be explained by the trends presented in Section~\ref{radprofile}. When the feedback efficiency becomes high, the accretion rate will be suppressed to ensure that the total energy output roughly balances the radiative cooling. Since the accretion rate becomes low, the black hole growth will become slower, which makes the result deviate from the observed $M_{\star}-M_{\rm{BH}}$ relation.

\subsection{The accretion fraction}\label{accf}

\begin{figure}
 \includegraphics[width=\linewidth]{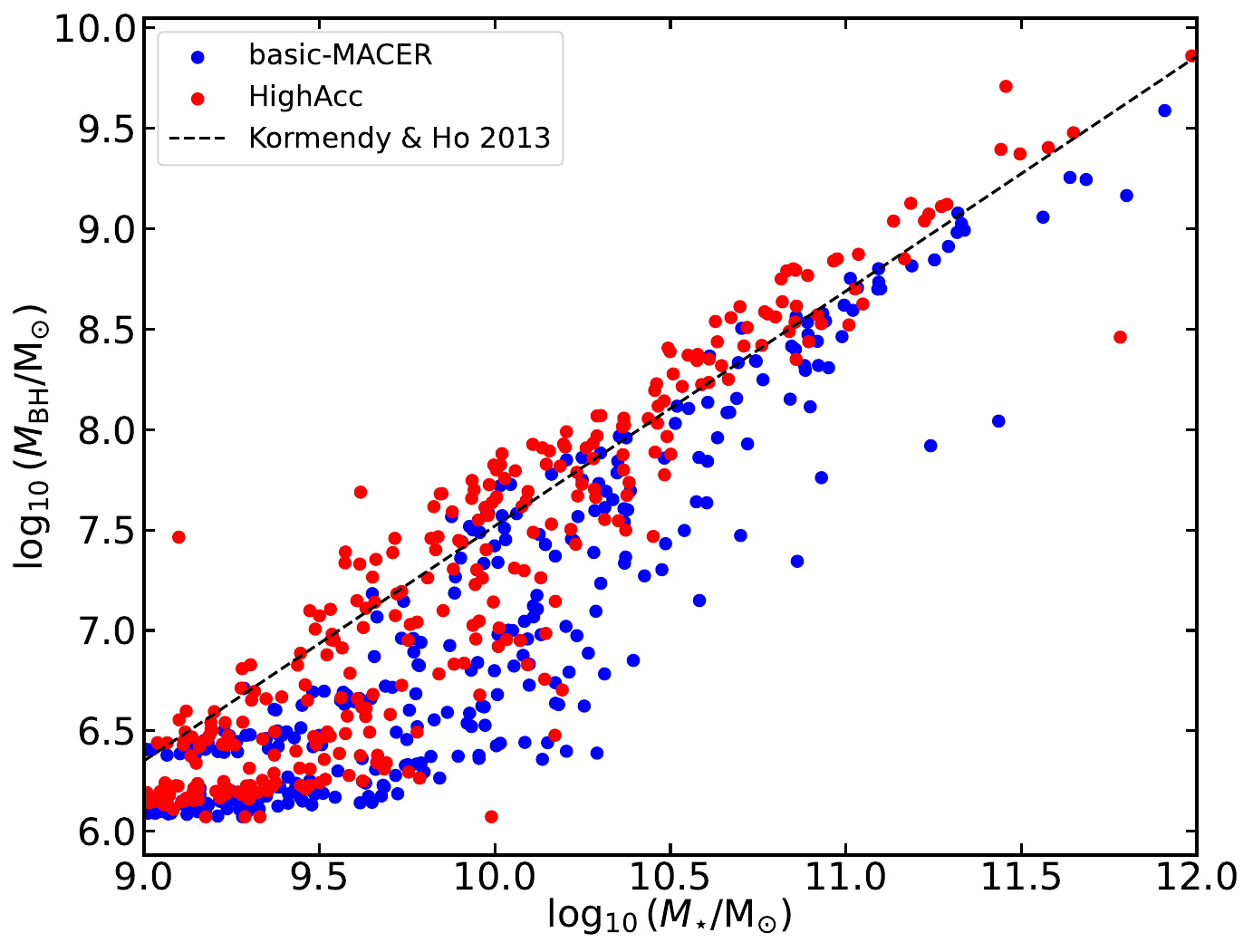}
 \caption{Comparison of the black hole mass versus stellar mass   relation produced by the  basic-MACER and HighAcc models at $z=0$. Here HighAcc is a model variant where  the basic-MACER is evolved with an accretion fraction equal to 0.9. The figure demonstrates that adopting a higher accretion fraction can make the low-mass BH grow considerably faster and reach the observed stellar mass-black hole mass relation already in lower mass galaxies.}
 \label{fig:ps4mbhmb}
\end{figure}

In basic-MACER model, a fraction of the material accreted from large radii will be converted to an outflow. To better understand the effects of the accretion fraction, we compare the results of the basic-MACER and Highacc simulations. In Fig.~\ref{fig:ps4mbhmb}, we show the $M_{\star}-M_{\rm{BH}}$ relation in the HighAcc and basic-MACER runs. The figure shows that the black hole masses in the HighAcc simulations are systematically larger than in basic-MACER, for all mass ranges. The discrepancy is larger at the low mass end than at the high mass end. 

\begin{figure}
 \includegraphics[width=\linewidth]{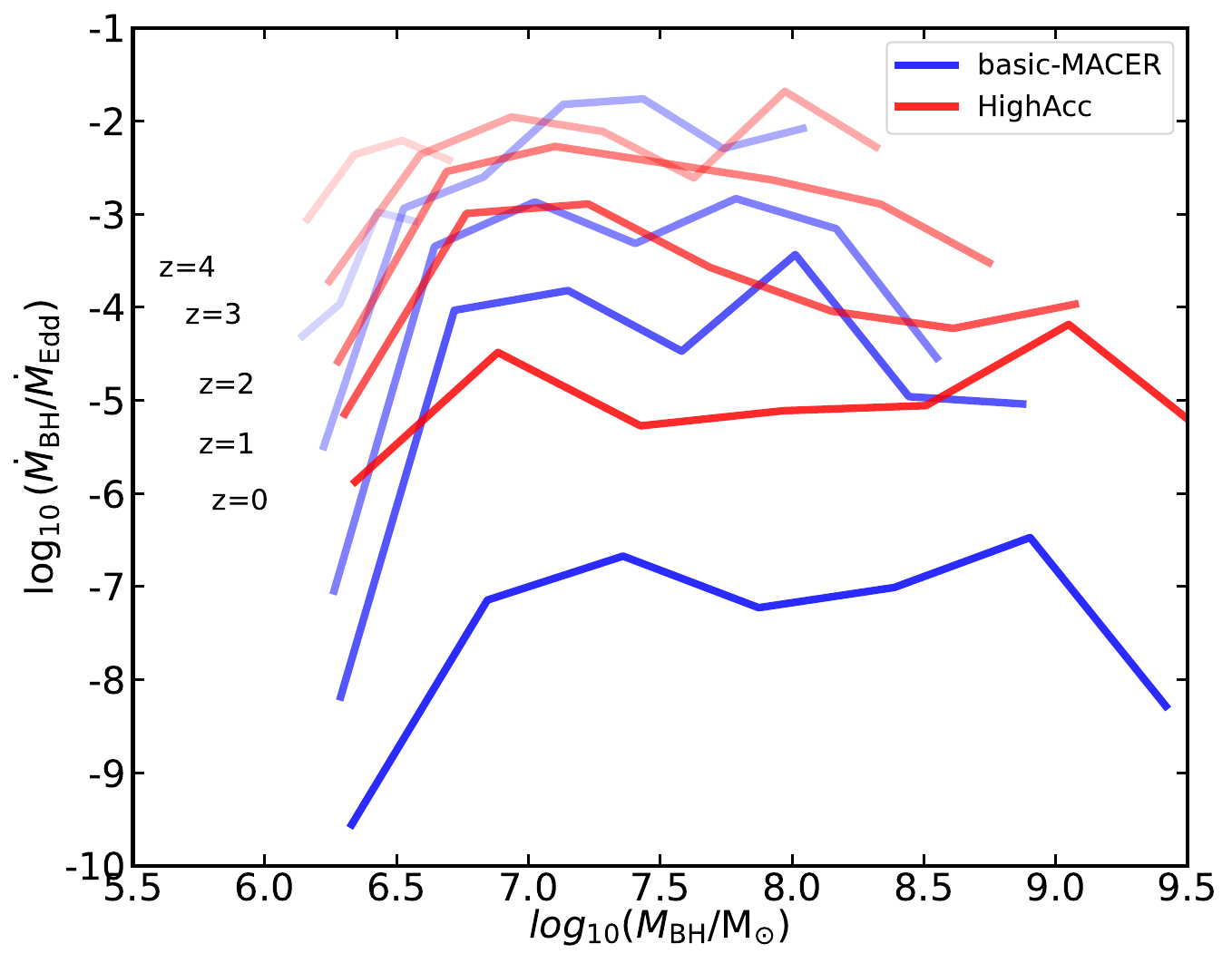}
 \caption{Eddington ratio as a function of black hole mass at different redshifts in the basic-MACER and HighAcc simulations. The figure shows that the actual BH growth rate  in the fiducial basic-MACER simulation is much lower than in the HighAcc model variant, especially  at $z<3$.}
 \label{fig:ps4mdotbh}
\end{figure}

We further plot the median BHAR ($\equiv f_{\rm{acc}}\dot{M}_{{\rm Bondi}}$) as a function of black hole mass at different redshifts for the basic-MACER and HighAcc models in Fig.~\ref{fig:ps4mdotbh}. From the figure, we can see that the BHAR is significantly lower in basic-MACER from $z=3$ to $z=0$, and the discrepancy becomes larger at low redshift. This difference in the BHAR implies that, over cosmic time, black holes in the basic-MACER model experience systematically slower mass growth. As a result, the cumulative mass difference between the two models becomes more pronounced, particularly in the low-mass regime.

\section{Summary}\label{sec:conclusion}

The AGN feedback model in MACER \citep{yuan18} features a state-of-the-art description of AGN physics, including a detailed calculation of the black hole accretion rate and the AGN outputs (from winds, jets, and radiation) as a function of accretion rate. In this work, we tried to implement key parts of this feedback model into cosmological simulions by modifying the model used in the IllustrisTNG project. Since the high spatial resolution of the original MACER project cannot easily be realized in cosmological simulations, we for the moment needed to be content with an approximate numerical implementation that is still similar to TNG. We name this modified  MACER AGN-feedback model the ``basic-MACER'' model. 

In this scenario, we keep the original estimation of the black hole mass accretion rate in terms of the Bondi accretion rate, as adopted in TNG. For the other aspects, the basic-MACER model makes the following three modifications compared to TNG:
\begin{enumerate}
\item The basic-MACER model changes the form in which feedback energy  is released in the quasar mode. Instead of using thermal energy injection, the basic-MACER model directly injects the kinetic energy of an AGN wind, which is likely the dominant output in this mode. The injected wind has both momentum and energy interaction with the ISM. The numerical implementation is the same as the radio mode in TNG. The parameters use results from observations of UFOs.

\item The basic-MACER model significantly changes the parameters of the AGN feedback in the radio mode, including the feedback efficiency, which is determined by the properties of winds and jets launched from the hot accretion flow of the black hole, and the fraction of the accreted gas estimated by the Bondi accretion formula that finally reaches the black hole horizon. The values of these parameters are obtained from general relativistic MHD simulations of black hole hot accretion flows \citep[e.g.,][]{yuan15}. 

\item The basic-MACER model changes the critical accretion rate that sets the transition between the quasar and radio modes. In the TNG model, this critical accretion rate changes with the black hole mass. In contrast, in the basic-MACER model, the critical accretion rate is set to be $0.01\,\dot{M}_{\rm Edd}$, which corresponds to $0.01\,L_{\rm Edd}$. This is suggested by  observations of black hole X-ray binaries and by black hole accretion theory \citep{yuan14}, 

\end{enumerate}

Since an idealized elliptical galaxy has been well simulated in the MACER framework, we perform low-resolution idealized elliptical galaxy simulations with the basic-MACER and TNG models as a first  example of studying the effects of the two feedback models. Through comparison of the results, we find that the simulated elliptical galaxy with the basic-MACER model has a higher SFR and BHAR compared to the  TNG model. This is because the TNG model has a higher feedback efficiency, resulting in a lower gas density at the central regions of the simulated elliptical galaxies. The density radial profile in our basic-MACER model is close to that predicted by the original MACER model.

We have then performed cosmological simulations in a periodic box with $50\,h^{-1}{\rm Mpc}$ on a side, using the basic-MACER and TNG models, and we compared the results. The main findings are summarized as follows:

\begin{enumerate}

\item {The basic-MACER model reproduces the star-forming main sequence for sub-$L^{\star}$ galaxies in good agreement with both TNG and observations. However, basic-MACER struggles to suppress star formation in massive galaxies compared to TNG, resulting in weaker sSFR bimodality and colour bimodality than observed.} This may be explained by results of the idealized elliptical galaxy simulations: the basic-MACER model produces massive galaxies with higher central gas density, which furthermore leads to a higher SFR.

\item Both the basic-MACER and TNG models can produce the observed $M_{\star}-M_{\rm BH}$ relation at $z=0$ for stellar masses $M_{\star}>10^{10.5}\,{\rm M_{\odot}}$. However, there are more low-mass galaxies with smaller BH for $M_{\star}<10^{10.5}\,{\rm M_{\odot}}$ in the basic-MACER model, although a mild resolution dependence remains in this low-mass regime.

\item The Bondi accretion rates of the massive BHs in the basic-MACER model are systematically higher than in the TNG model, which can also be explained by the fact that the basic-MACER model produces a higher central gas density. However, since only a small fraction of the gas finally supports the BH growth in the basic-MACER model, the BH mass will not grow more dramatically compared to the TNG model.

\item The gas fraction in the basic-MACER model shows a smooth increase with increasing halo mass, while the TNG model exhibits a strong discontinuity at $M_{\rm halo}\sim10^{12}\,{\rm M_{\odot}}$. This is because the feedback mode in the TNG model switches from quasar mode to radio mode around this halo mass. As a result, the gas fraction varies  strongly at this halo mass scale.

\end{enumerate}

We have also performed many ``hybrid'' experiments between the TNG and basic-MACER models to better understand the various parameter dependencies in the AGN feedback models, and obtained the following results by exchanging parts of the TNG and basic-MACER models:

\begin{enumerate}
\item The hydrid model with the quasar mode of the TNG model and the radio mode of the basic-MACER model reproduces a better bimodality of the galaxy colour compared to the basic-MACER model, and the $M_{\star}-M_{\rm BH}$ relation is also similar to the basic-MACER model despite that the BH mass has a large jump at $M_{\star}\sim10^{10}\,{\rm M_{\odot}}$. 

\item Changing the feedback efficiency  with the BH mass in the quasar mode of the basic-MACER model, we can reproduce the observed galaxy population. The model with  a constant feedback efficiency strongly suppresses the BH growth.

\item The statistical results of the simulated galaxy populations are not very sensitive to inclusion of the jet efficiency in the radio mode. However, if the feedback efficiency in the radio mode is too high, it also affects the BH growth, causing the simulated $M_{\star}-M_{\rm BH}$ relation to deviate from observations. We note, however, that our jet model does not include an explicit treatment of black hole spin evolution. Recent cosmological sub-grid models that incorporate spin-dependent jet efficiencies \citep[e.g.,][]{husko2025} suggest that spin-regulated jets may play an important role at low accretion rates since the jet efficiency can reach several hundred percent in extreme states with near-maximal black hole spin and magnetically-arrested accretion flows, potentially enhancing the impact of radio-mode feedback.

\item The accretion fraction parameter introduced in the basic-MACER model controls the BH growth, and thus affects the $M_{\star}-M_{\rm BH}$ relation significantly. Compared to  massive BHs, the low mass BHs are more sensitive to the value of this parameter.

\end{enumerate}

Recent studies have explored a wide range of AGN feedback prescriptions in cosmological simulations, including kinetic jet implementations \citep[e.g.,][]{dubois2021}, alternative wind driving schemes \citep[e.g.,][]{farcy2025}, and sub-grid models that link jet efficiencies to black hole accretion states and spin evolution \citep[e.g.,][]{husko2025}. Despite substantial differences in the physical implementations, a recurring theme in these works is that black holes tend to self-regulate: different feedback implementations mainly change the black hole mass at which self-regulation sets in, but finally lead to similar statistical properties of galaxies.

Our results are consistent with this picture. In particular, similar to \citet{husko2025}, we find that modifying the efficiency parameters mainly affects black hole growth and the quenching efficiency of massive galaxies, whereas the overall SFRD, GSMF and $M_\star$–$M_{\rm BH}$ relation remain comparatively robust within our cosmological volume. This suggests that, at the level of galaxy population statistics, the dominant role of AGN feedback is not the specific feedback channel itself, but the establishment of a self-regulated coupling between black hole growth and the surrounding gas.

In this work, the jet feedback is implemented using a simplified phenomenological prescription, where the jet efficiency is treated as a parameter and does not explicitly depend on black hole spin or the magnetic flux threading the horizon. As a result, our radio-mode feedback does not capture the potentially strong spin-dependent effects suggested by recent GRMHD simulations and cosmological sub-grid models with spin evolution \citep[e.g.,][]{husko2025}. These studies indicate that, in extreme states of the black hole and the accretion flow, jet efficiencies can exceed unity, potentially enhancing the impact of radio-mode feedback. Incorporating such physics self-consistently in cosmological simulations is beyond the scope of the present work, but will be an important direction for future improvements of AGN feedback models.

In summary, the basic-MACER model can reproduce the galaxy properties of the TNG model under the same numerical setup, such as the SFRD, GSMF and the $M_{\star}-M_{\rm BH}$ relation. At the same time, several differences with respect to TNG are found, including a lower central gas density in massive ellipticals, a better reproduction of low-mass black holes in low-mass galaxies, and a milder quenching of massive galaxies. On the other hand, some issues such as obtaining a bimodality of the galaxy colours and a sufficiently strong quenching of massive galaxies are not readily resolved in the basic-MACER model. This could be because of the still oversimplified treatment of the black hole accretion rate and the energy deposition of the AGN outputs in the ISM. It is crucial in future work to improve on these two issues in cosmological simulations.
\section*{Acknowledgements}
The authors thank the anonymous reviewer for helpful comments. B.Z. and F.Y. are supported by Natural Science Foundation of China (grants No. 12133008, 12192220, 12192223, and 12361161601), China Manned Space Project (grants CMS-CSST-2025-A08 and CMS-CSST-2025-A10), and the National SKA Program of China (No. 2025SKA0130100). The authors acknowledge support by a Max Planck Partner group between the Shanghai Astronomical Observatory and the Max Planck Institute for Astrophysics (MPA). Computations were performed on the Freya compute cluster of MPA, operated by the Max Planck Computing and Data Facility.

\section*{Data Availability}
The data of this study are available from the corresponding author upon reasonable request.



\bibliographystyle{mnras}




\bibliography{ref}

\bsp	
\label{lastpage}
\end{document}